\pdfoutput=1
\documentclass[journal,twoside,web]{ieeecolor}

\usepackage{tmi}

\def\BibTeX{{\rm B\kern-.05em{\sc i\kern-.025em b}\kern-.08em
    T\kern-.1667em\lower.7ex\hbox{E}\kern-.125emX}}

\makeatletter
\let\NAT@parse\undefined
\makeatother

\usepackage{blindtext} %
\usepackage{mwe}  %

\usepackage[T1]{fontenc}  %
\usepackage[utf8]{inputenc}  %

\usepackage{newtxmath}
\usepackage{bm}  %
\DeclareMathAlphabet{\mathcal}{OMS}{cmsy}{m}{n}
\DeclareSymbolFont{AMSb}{U}{msb}{m}{n}
\DeclareSymbolFontAlphabet{\mathbb}{AMSb}

\usepackage{csquotes} %
\usepackage{textcomp}  %
\usepackage{siunitx}  %
\DeclareSIUnit{\fps}{fps}
\usepackage{newunicodechar} %
\usepackage{pifont} %
\usepackage{bbding} %
\usepackage{xfrac}
\usepackage[super]{nth}

\usepackage[final, expansion=alltext, protrusion=alltext-nott]{microtype}
\DisableLigatures{encoding = T1, family = tt*}

\usepackage[english]{babel}
\addto\captionsenglish{%
}

\usepackage[style=base]{caption}  %
\usepackage{subcaption}
\DeclareCaptionLabelSeparator{ieeefigsep}{.\quad}
\DeclareCaptionLabelSeparator{ieeetmifigsep}{\textcolor{mblue}{.}\quad}
\usepackage{mathtools}  %
\usepackage{cases}

\usepackage{algorithm}
\usepackage{algpseudocode}
\usepackage[zerostyle=a,straightquotes]{newtxtt}

\usepackage{booktabs}
\usepackage{multirow}
\usepackage{threeparttable}
\makeatletter
\def\TPT@doparanotes{\par
   \prevdepth\z@ \TPT@hsize
   \TPTnoteSettings
   \parindent\z@ \pretolerance 8
   \linepenalty 200
   \renewcommand\item[1][]{\relax\ifhmode \begingroup
       \unskip
       \advance\hsize 10em %
       \penalty -45 \hskip\z@\@plus\hsize \penalty-19
       \hskip .15\hsize \penalty 9999 \hskip-.15\hsize
       \hskip .01\hsize\@plus-\hsize\@minus.01\hsize
       \hskip .5em\@plus .3em
      \endgroup\fi
      \tnote{##1}\ignorespaces}%
   \let\TPToverlap\relax
   \def\endtablenotes{\par}%
}
\makeatother
\renewcommand{\TPTnoteSettings}{%
    \setlength\leftmargin{1.5em}%
    \setlength\labelwidth{.5em}%
    \setlength\labelsep{0pt}%
    \rightskip\tabcolsep \leftskip\tabcolsep
}
\usepackage{tabularx}
\newcolumntype{L}[1]{>{\hsize=\if!#1!1\else#1\fi\hsize\raggedright\arraybackslash}X}
\newcolumntype{R}[1]{>{\hsize=\if!#1!1\else#1\fi\hsize\raggedleft\arraybackslash}X}
\newcolumntype{C}[1]{>{\hsize=\if!#1!1\else#1\fi\hsize\centering\arraybackslash}X}
\usepackage{glossaries-extra}
\setabbreviationstyle[acronym]{long-short}  %

\usepackage{calc}
\usepackage{xifthen} %
\usepackage{etoolbox} %
\usepackage{xparse} %

\usepackage{soul}
\usepackage{printlen}

\usepackage{cite}

\usepackage{url}
\usepackage[bookmarks=false, hidelinks]{hyperref}
\usepackage[nameinlink, capitalize]{cleveref} %
\crefname{table}{Table}{Tables} %
\crefname{figure}{Fig.\@}{Fig.\@} %
\Crefname{figure}{Fig.\@}{Fig.\@} %
\crefname{section}{Section}{Sections} %
\Crefname{section}{Section}{Sections} %
\crefname{equation}{}{} %
\Crefname{equation}{Equation}{Equations} %
\crefname{algorithm}{Algorithm}{Algorithms} %
\Crefname{algorithm}{Algorithm}{Algorithms} %

\let\originalleft\left
\let\originalright\right
\renewcommand{\left}{\mathopen{}\mathclose\bgroup\originalleft}
\renewcommand{\right}{\aftergroup\egroup\originalright}
\DeclarePairedDelimiter{\parens}{\lparen}{\rparen}
\DeclarePairedDelimiter{\braces}{\lbrace}{\rbrace}

\DeclarePairedDelimiterXPP\onenorm[1]{}{\lVert}{\rVert}{_{1}}{#1}
\DeclarePairedDelimiterXPP\twonorm[1]{}{\lVert}{\rVert}{_{2}}{#1}
\DeclarePairedDelimiterXPP\fronorm[1]{}{\lVert}{\rVert}{_{\text{F}}}{#1}
\DeclarePairedDelimiterXPP\infnorm[1]{}{\lVert}{\rVert}{_{\infty}}{#1}
\DeclarePairedDelimiterXPP\pnorm[1]{}{\lVert}{\rVert}{_{p}}{#1}
\DeclarePairedDelimiterXPP\tvnorm[1]{}{\lVert}{\rVert}{_{\text{TV}}}{#1}
\DeclarePairedDelimiterXPP\maxnorm[1]{}{\lVert}{\rVert}{_{\text{max}}}{#1}
\DeclarePairedDelimiterXPP\nuclearnorm[1]{}{\lVert}{\rVert}{_{\ast}}{#1}

\newcommand*{\set}[1]{\mathbb{#1}}

\NewDocumentCommand{\realnumbers}{}{\set{R}}

\NewDocumentCommand{\ellipsis}{}{\dots}
\let\setroster\braces

\newcommand*{\mat}[1]{\bm{#1}}
\renewcommand*{\vec}[1]{\bm{#1}}

\newcommand*{\estimate}[1]{\hat{#1}}

\newcommand*{\argdot}{\,\cdot\,}

\newcommand*{\wavelength}{\lambda}

\newcommand{\aperture}{L}

\newcommand*{\imageletter}{x}

\newcommand*{\imvec}{\vec{\imageletter}}

\newcommand*{\imvecapprox}{\tilde{\imvec}}

\newcommand*{\imvecdim}{n}
\newcommand*{\imvecdom}{\realnumbers^{\imvecdim}}

\newcommand*{\rawdataletter}{y}

\newcommand*{\rdvec}{\vec{\rawdataletter}}

\newcommand*{\rdvecdim}{m}
\newcommand*{\rdvecdom}{\realnumbers^{\rdvecdim}}
\newcommand*{\rdvecdef}{\rdvec\in\rdvecdom}

\newcommand*{\dasopletter}{D}
\newcommand*{\dasopmat}{\mat{\dasopletter}}
\newcommand*{\dasopmatdef}{\dasopmat\colon\rdvecdom\to\imvecdom}

\newcommand*{\cnnparams}{\vec{\theta}}
\newcommand*{\cnnletter}{f}
\newcommand*{\cnnsymb}{\vec{\cnnletter}_{\cnnparams}}
\newcommand*{\cnn}{\cnnsymb}

\newcommand*{\cnndef}{\cnn\colon\imvecdom\to\imvecdom}

\newcommand*{\dsetsize}{l}

\newcommand{\dispvector}{\vec{u}}

\newcommand*{\pivframe}{\mat{S}}

\newcommand*{\cpwcanglenumberref}{\cpwcanglenumber^{\text{ref}}}

\newcommand*{\cpwcanglenumber}{N_{a}}
\newcommand*{\cpwcangle}{\beta}

\newcommand*{\cpwcanglespacing}{\Delta\cpwcangle}
\newcommand*{\angleind}{\cpwcangle_{n}}
\newcommand*{\cpwcanglemaxindex}{M}
\newcommand*{\cpwcmethodname}[1]{\mbox{\glsxtrshort{cpwc}-#1}}
\newcommand*{\fnumber}{F_{\#}}

\newcommand*{\ndim}[1]{\mbox{#1-D}}  %

\NewDocumentCommand{\margintodo}{ m o }{%
    \marginpar{\colorbox{yellow}{\parbox{\marginparwidth-\marginparsep}{\raggedright\scriptsize \textcolor{black}{#1}}}}%
    \IfValueT{#2}{\hl{#2}}%
}
\NewDocumentCommand{\inlinetodo}{ m o }{%
    \IfValueT{#2}{\hl{#2}}\raisebox{+0.85ex}{\tiny\hl{[#1]}}%
}

\newcommand{\invalidtabentry}{\( \times \)}

\newcommand{\sprctext}{ten} %

\newcommand*{\vsxtxfreq}{\SI{5.208}{\mega\hertz}}
\newcommand*{\vsxrxfreq}{\SI{20.833}{\mega\hertz}}
\newcommand*{\vsxprf}{\SI{9}{\kilo\hertz}}

\newcommand*{\geaperture}{\SI{43.93}{\milli\meter}}
\newcommand*{\gecenterfreq}{\SI{5.3}{\mega\hertz}}
\newcommand*{\gebandwidth}{\SI{75}{\percent}}
\newcommand*{\geelemnumber}{\num{192}}
\newcommand*{\geelemwidth}{\SI{207}{\micro\meter}}
\newcommand*{\geelemheight}{\SI{6}{\milli\meter}}
\newcommand*{\geelevationfocus}{\SI{28}{\milli\meter}}
\newcommand*{\gepitch}{\SI{230}{\micro\meter}}

\newcommand*{\imagesampling}{\wavelength{} / 4 \times{} \wavelength{} / 8}
\newcommand*{\envsampling}{\wavelength{} / 4 \times{} \wavelength{} / 4}
\newcommand*{\numimdim}{\num{596 x 1600}}
\newcommand*{\numenvdim}{\num{596 x 800}}

\newcommand{\sectitlenamenumexp}{Numerical Experiment}
\newcommand*{\zylradius}{\SI{6.86}{\milli\meter}}
\newcommand*{\zylheight}{\SI{1.0}{\milli\meter}}
\newcommand*{\zylechoa}{\SI{20}{\decibel}}
\newcommand*{\zylechob}{\SI{-20}{\decibel}}
\newcommand*{\zylechoc}{\SI{-20}{\decibel}}
\newcommand*{\zylechod}{\SI{0}{\decibel}}
\newcommand*{\disprangetot}{\SIrange{3.3}{600}{\micro\meter}}
\newcommand*{\disprangelarge}{\SIrange{33}{600}{\micro\meter}}
\newcommand*{\disprangesmall}{\SIrange{3.3}{60}{\micro\meter}}
\newcommand*{\zylradcropp}{\SI{6.5}{\milli\meter}}
\newcommand*{\zylborder}{\SI{0.36}{\milli\meter}}
\newcommand*{\numframessim}{\num{50}}
\newcommand{\edtnumphtcylAlbl}{\text{A}}
\newcommand{\edtnumphtcylBlbl}{\text{B}}
\newcommand{\edtnumphtcylClbl}{\text{C}}
\newcommand{\edtnumphtcylDlbl}{\text{D}}

\newcommand{\sectitlenamephysexp}{\textit{In Vivo} Experiment}
\newcommand*{\expfreq}{\SI{10}{\hertz}}
\newcommand*{\exproisize}{\SI{5 x 5}{\milli\meter}}

\newcommand*{\pyusurl}{\url{https://gitlab.com/pyus/pyus}}

\newglossaryentry{python}{%
    name={Python},
    description={interpreted, high-level, general-purpose programming language}
}
\newglossaryentry{matlab}{%
    name={MATLAB},
    description={multi-paradigm numerical computing environment and proprietary programming language developed by MathWorks}
}
\newglossaryentry{tensorflow}{%
   name={TensorFlow},
   description={An end-to-end open source machine learning platform}
}
\newglossaryentry{tensorrt}{%
   name={TensorRT},
   description={NVIDIA TensorRT™ is a platform for high-performance deep learning inference.}
}
\newglossaryentry{mu-law}{%
    name={\textmu-law},
    description={Comanding algorithm}
}
\newglossaryentry{ge}{%
    name={General Electric Company},
    description={TODO}
}

\newglossaryentry{ge-healthcare}{%
    name={GE Healthcare},
    description={TODO}
}
\newcommand{\gehealthcareaddress}{Chicago, Illinois, USA}
\newglossaryentry{ge9ld}{%
    name={9L-D},
    description={GE Linear transducer}
}
\newglossaryentry{bmode}{%
    name={B-mode},
    description={Brightness mode}
}
\newglossaryentry{verasonics}{%
    name={Verasonics},
    description={TODO}
}
\newglossaryentry{vantage256}{%
    name={Vantage~256},
    description={TODO}
}
\newcommand{\verasconicsaddress}{Kirkland, WA, USA}
\newglossaryentry{cirs}{%
    name={CIRS},
    description={Multi-purpose multi-tissue ultrasound phantom}
}

\newglossaryentry{cirs-040gse}{%
    name={\gls{cirs} model~040GSE},
    description={Multi-purpose multi-tissue ultrasound phantom}
}
\newglossaryentry{cirs-054gs}{%
    name={\gls{cirs} model~054GS},
    description={General purpose ultrasound phantom}
}
\newglossaryentry{pyus}{%
    name={PyUS},
    description={TODO}
}
\newglossaryentry{fieldii}{%
    name={Field~II},
    description={TODO}
}
\newglossaryentry{pivlab}{%
    name={PIVlab},
    description={TODO}
}
\newglossaryentry{x-ray}{%
    name={X-ray},
    description={X-radiation or Röntgen radiation}
}
\newglossaryentry{nvidia}{%
    name={NVIDIA},
    description={American technology company incorporated in Delaware and based in Santa Clara, California. It designs graphics processing units (GPUs) for the gaming and professional markets, as well as system on a chip units (SoCs) for the mobile computing and automotive market.}
}
\newglossaryentry{1080ti}{
    name={NVIDIA GeForce GTX 1080~Ti},
    description={Specific GPU from NVIDIA (Pascal architecture)}
}
\newglossaryentry{mx150}{
    name={NVIDIA GeForce MX~150},
    description={Specific GPU from NVIDIA (Pascal architecture)}
}
\newglossaryentry{titanv}{
    name={NVIDIA TITAN~V},
    description={Specific GPU from NVIDIA (Volta architecture)}
}
\newglossaryentry{v100}{
    name={NVIDIA Tesla V100},
    description={Specific GPU from NVIDIA (Volta architecture)}
}
\newglossaryentry{unet}{
    name={U-Net},
    description={TODO}
}
\newglossaryentry{resnet}{
    name={ResNet},
    description={Typical neural network architecture}
}
\newglossaryentry{bspline}{
    name={B-spline},
    description={TODO}
}
\newglossaryentry{etc}{
    name={etc.},
    description={et cetera},
}
\newglossaryentry{ie}{
    name={i.e.}, description={id est}, %
}
\newglossaryentry{eg}{
    name={e.g.}, description={exempli gratia}, %
}
\newglossaryentry{etal}{
    name={\textit{et al.}}, description={et alii}, %
}
\newglossaryentry{ibid}{
    name={\textit{ibid.}},
    description={ibidem}, %
}
\newglossaryentry{vs}{
    name={vs.}, description={TODO}, %
}
\newglossaryentry{wrt}{
    name={w.r.t.}, description={TODO}, %
}
\newglossaryentry{esp}{
    name={esp.}, description={TODO}, %
}
\newglossaryentry{invitro}{
    name={\textit{in vitro}}, description={TODO}, %
}
\newglossaryentry{invivo}{
    name={\textit{in vivo}}, description={TODO}, %
}
\newglossaryentry{exvivo}{
    name={\textit{ex vivo}}, description={TODO}, %
}
\newglossaryentry{insitu}{
    name={\textit{in situ}}, description={TODO}, %
}
\newglossaryentry{insilico}{
    name={\textit{in silico}}, description={TODO}, %
}
\newglossaryentry{interalia}{
    name={\textit{inter alia}}, description={TODO}, %
}
\newglossaryentry{intoto}{
    name={\textit{in toto}}, description={TODO}, %
}
\newglossaryentry{apriori}{
    name={\textit{a priori}}, description={TODO}, %
}
\newglossaryentry{aposteriori}{
    name={\textit{a posteriori}}, description={TODO}, %
}
\newglossaryentry{alas}{
    name={\textit{alas}}, description={TODO}, %
}
\newglossaryentry{defacto}{
    name={de facto}, description={TODO}, %
}

\newacronym{nn}{NN}{neural network}
\newacronym{dnn}{DNN}{deep neural network}
\newacronym{cnn}{CNN}{convolutional neural network}
\newacronym{relu}{ReLU}{rectified linear unit}
\newacronym{conv-layer}{CL}{convolutional layer}
\newacronym{std-conv-block}{FCB}{fully convolutional block}
\newacronym{res-conv-block}{RCB}{residual convolutional block}
\newacronym{mse}{MSE}{mean squared error}
\newacronym{mae}{MAE}{mean absolute error}
\newacronym{mslae}{MSLAE}{mean signed logarithmic absolute error}
\newacronym{mmuae}{MMUAE}{mean \textmu-law absolute error}
\newacronym{roi}{ROI}{region of interest}
\newacronym{voi}{VOI}{volume of interest}
\newacronym{fps}{FPS}{frames per second}
\newacronym{cpwc}{CPWC}{coherent plane wave compounding}
\newacronym{cr}{CR}{constrast ratio}
\newacronym{snr}{SNR}{signal-to-noise ratio}
\newacronym{cnr}{CNR}{contrast-to-noise ratio}
\newacronym{gcnr}{GCNR}{generalized contrast-to-noise ratio}
\newacronym{fwhm}{FWHM}{full width at half maximum}
\newacronym{ssim}{SSIM}{structural similarity}
\newacronym{psnr}{PSNR}{peak signal-to-noise ratio}
\newacronym{dra}{DRA}{dynamic range alteration}
\newacronym{epe}{EPE}{endpoint error}
\newacronym{mepe}{MEPE}{mean \glsxtrlong{epe}}
\newacronym{repe}{REPE}{relative \glsxtrlong{epe}}
\newacronym{mrepe}{MREPE}{mean \glsxtrlong{repe}}
\newacronym{rve}{RVE}{ratio of valid estimates}
\newacronym{mri}{MRI}{magnetic resonance imaging}
\newacronym{ct}{CT}{computed tomography}
\newacronym{hdr}{HDR}{high dynamic range}
\newacronym{nurbs}{NURBS}{non-uniform rational B-spline}
\newacronym{bp}{BP}{backprojection}
\newacronym{fbp}{FBP}{filtered backprojection}
\newacronym{us}{US}{ultrasound}
\newacronym{das}{DAS}{delay-and-sum}
\newacronym{trf}{TRF}{tissue reflectivity function}
\newacronym{pw}{PW}{plane wave}
\newacronym{sa}{SA}{synthetic aperture}
\newacronym{dw}{DW}{diverging wave}
\newacronym{rf}{RF}{radio frequency}
\newacronym{prf}{PRF}{pulse repetition frequency}
\newacronym{tgc}{TGC}{time gain compensation}
\newacronym{lq}{LQ}{low-quality}
\newacronym{hq}{HQ}{high-quality}
\newacronym{uq}{UQ}{ultra-high-quality}
\newacronym{picmus}{PICMUS}{plane-wave imaging challenge in medical ultrasound}
\newacronym{sir}{SIR}{spatial impulse response}
\newacronym{psf}{PSF}{point spread function}
\newacronym{iq}{IQ}{in-phase quadrature}
\newacronym{gl}{GL}{grating lobe}
\newacronym{sl}{SL}{side lobe}
\newacronym{ew}{EW}{edge wave}
\newacronym{tx}{Tx}{transmit}
\newacronym{rx}{Rx}{receive}
\newacronym{mv}{MV}{minimum variance}
\newacronym{piv}{PIV}{particle image velocimetry}
\newacronym{zncc}{ZNCC}{zero-normalized cross-correlation}
\newacronym{epfl}{EPFL}{École polytechnique fédérale de Lausanne}
\newacronym{lts5}{LTS5}{Signal Processing Laboratory 5}
\newacronym{chuv}{CHUV}{University Hospital Center}
\newacronym{unil}{UNIL}{University of Lausanne}
\newacronym{cibm}{CIBM}{Center for Biomedical Imaging}
\newacronym{gpu}{GPU}{graphics processing unit}
\newacronym{ieee}{IEEE}{Institute of Electrical and Electronics Engineers}
\newacronym{ius}{IUS}{International Ultrasonic Symposium}
\newacronym{sa-vfi}{SA-VFI}{synthetic aperture vector flow imaging}

\title{%
    CNN-Based Ultrasound Image Reconstruction for
    Ultrafast Displacement Tracking
}

\newcommand*{\correspondingemail}{dimitris.perdios@epfl.ch}
\newcommand*{\correspondingauthor}{Dimitris Perdios}

\author{%
    Dimitris~Perdios,~\IEEEmembership{Student~Member,~IEEE,}
    Manuel~Vonlanthen,
    Florian~Martinez,~\IEEEmembership{Member,~IEEE,}
    Marcel~Arditi,~\IEEEmembership{Senior~Member,~IEEE,}
    and
    Jean-Philippe~Thiran,~\IEEEmembership{Senior~Member,~IEEE}%
    \thanks{%
        This work was supported in part by the Swiss National Science Foundation
        under Grant 205320\_175974 and Grant 206021\_170758.
        \textit{%
            (Dimitris Perdios and Manuel Vonlanthen
            contributed equally to this work.)
        }
        \textit{%
            (Corresponding author:
            \href{mailto:\correspondingemail}{\correspondingauthor}.)
        }
    }
    \thanks{%
        D.~Perdios, M.~Vonlanthen, F.~Martinez, M.~Arditi, and J.-Ph.~Thiran
        are with the
        \glsxtrfull{lts5},
        \glsxtrfull{epfl},
        1015 Lausanne,
        Switzerland
        (email: \href{mailto:\correspondingemail}{\correspondingemail}).
    }
    \thanks{%
        J.-Ph.~Thiran
        is also with the
        Department of Radiology,
        \glsxtrfull{chuv}
        and
        \glsxtrfull{unil},
        1011 Lausanne,
        Switzerland,
        and with the
        \glsxtrfull{cibm},
        1015 Lausanne,
        Switzerland.
    }
}

\begin{document}

\maketitle

\glsresetall{}
\begin{abstract}

Thanks to its capability of acquiring full-view frames at multiple kilohertz,
ultrafast \glsxtrlong{us} imaging unlocked the analysis
of rapidly changing physical phenomena in the human body,
with pioneering applications such as ultrasensitive
flow imaging in the cardiovascular system or shear-wave elastography.
The accuracy achievable with these motion estimation techniques
is strongly contingent upon two
contradictory requirements:
a high quality of consecutive frames and a high frame rate.
Indeed,
the image quality can usually be improved by increasing the number of
steered ultrafast acquisitions,
but at the expense of a reduced frame rate and possible motion artifacts.
To achieve accurate motion estimation at uncompromised frame rates
and immune to motion artifacts,
the proposed approach relies on single ultrafast acquisitions
to reconstruct high-quality frames
and on only two consecutive frames to obtain \ndim{2} displacement estimates.
To this end,
we deployed a \glsxtrlong{cnn}-based image reconstruction method
combined with a speckle tracking algorithm based on cross-correlation.
Numerical and \gls{invivo} experiments,
conducted in the context of plane-wave imaging,
demonstrate that the proposed approach is capable of estimating displacements
in regions where the presence of \glsxtrlong{sl} and \glsxtrlong{gl} artifacts
prevents any displacement estimation with a state-of-the-art technique
that relies on conventional \glsxtrlong{das} beamforming.
The proposed approach may therefore unlock the full potential of
ultrafast ultrasound,
in applications such as ultrasensitive cardiovascular motion
and flow analysis or shear-wave elastography.

\end{abstract}


\glsresetall{}

\section{Introduction}%
\label{sec:introduction}

\IEEEPARstart{U}{ltrafast} \gls{us} imaging enables reconstructing
full-view images from single acquisitions by insonifying
the entire field of view at once,
using unfocused transmit wavefronts
such as \glspl{pw} or \glspl{dw}~\cite{Tanter_UFFC_2014}.
\Glsxtrlong{us} images are then reconstructed from the received
echo signals using the well-known \gls{das} algorithm.
Ultrafast \gls{us} imaging thus breaks with the trade-off between field of view
and frame rate inherent to conventional transmit-focused line-by-line scanning.
This enables imaging large tissue regions at very high frame rates
of multiple kilohertz,
limited only by the round-trip propagation time of single acoustic waves.

Imaging large tissue regions at such high frame rates is necessary for studying
the most rapidly changing physical phenomena in the human body,
such as tracking the propagation of naturally occurring or externally induced
shear waves~\cite{Bercoff_UFFC_2004,Montaldo_UFFC_2009,Santos_UFFC_2019,Papadacci_UFFC_2014,Pernot_JACC_2011}.
In the cardiovascular system,
where a frame rate of several hundred hertz is needed for resolving tissue
motion and flow patterns accurately~\cite{Geyer_JASE_2010,Cikes_JACC_2014,Voigt_EHJCI_2015, Bercoff_UFFC_2011},
ultrafast imaging enables increased ensemble lengths,
improving the robustness and sensitivity of displacement estimates
significantly~\cite{Bercoff_UFFC_2011}.
Several breakthrough \gls{us} imaging modes based on motion estimation
within a large field of view rely on ultrafast \gls{us} imaging,
such as shear-wave elastography~\cite{Bercoff_UFFC_2004},
ultrasensitive flow imaging~\cite{Bercoff_UFFC_2011},
and functional \gls{us} neuroimaging~\cite{Mace_NMETH_2011}.

Because of the absence of transmit focusing,
images obtained from ultrafast acquisitions are of low quality,
suffering heavily from poor lateral resolution
and low contrast~\cite{Montaldo_UFFC_2009,Cheng_UFFC_2006,Denarie_TMI_2013,
Papadacci_UFFC_2014,Cikes_JACC_2014,Santos_UFFC_2019}.
Both effects are related to the \gls{psf} of ultrafast \gls{us} imaging systems,
characterized by a broader main lobe (lower lateral resolution)
and stronger diffraction artifacts (lower contrast)
caused by \glspl{sl}, \glspl{gl}, and \glspl{ew},
compared with conventional focused-\gls{us} imaging systems.
Naturally, low-quality images also limit the accuracy of
subsequent displacement estimation methods involved in ultrafast \gls{us}
imaging modes~\cite{Montaldo_UFFC_2009,Papadacci_UFFC_2014,Voigt_EHJCI_2015}.
The state-of-the-art solution for increasing the quality of
ultrafast \gls{us} imaging is coherent compounding,
where a series of low-quality images,
reconstructed from multiple,
differently steered,
unfocused wavefronts,
are coherently summed~\cite{Cheng_UFFC_2006,Montaldo_UFFC_2009}.
In~\cite{Montaldo_UFFC_2009},
an image quality surpassing state-of-the-art multi-focus imaging
was obtained by compounding \num{71} \gls{pw} acquisitions,
increasing the frame-rate by a factor of approximately seven.

However,
for analyzing motion at very high frame rates,
coherent compounding suffers from two considerable disadvantages.
Firstly,
the increase in image quality is directly linked to the number
of compounded acquisitions,
which in turn is limited by the minimum frame rate necessary to analyze
the underlying physical phenomenon of interest.
Secondly,
coherent compounding assumes,
similarly to line-by-line scanning,
that the region of interest is stationary
for the duration of an acquisition sequence used to reconstruct a single frame.
This assumption does not hold when imaging fast-moving tissue regions
or complex flows,
for which coherent compounding suffers from strong motion
artifacts~\cite{Denarie_TMI_2013,Poree_TMI_2016}.

The first issue is well exemplified in~\cite{Montaldo_UFFC_2009},
in which Montaldo~\gls{etal} demonstrated,
in the context of shear-wave elastography,
that the quality of estimated elasticity maps is directly linked
to the number of compounded acquisitions,
which in turn was limited to a maximum of twelve acquisitions
to ensure a minimum frame rate of \SI{1}{\kilo\hertz}.
In particular,
displacement estimation in highly heterogeneous tissue regions,
where the aforementioned diffraction artifacts were dominant,
was a major obstacle.
Issues due to diffraction artifacts hindering accurate displacement estimates
have been reported for several methods,
all of them suffering from the trade-off between image quality
and frame rate~\cite{Montaldo_UFFC_2009,Papadacci_UFFC_2014,Poree_TMI_2015}.

The occurrence of severe motion artifacts when compounding multiple acquisitions
of rapidly evolving physical phenomena
(inter-frame displacement close to the effective wavelength)
was discussed in~\cite{Denarie_TMI_2013,Poree_TMI_2016,Joos_UFFC_2018},
and motion compensation techniques were proposed to tackle this problem.
They consist of estimating inter-acquisition displacement,
using either conventional Doppler~\cite{Poree_TMI_2016,Joos_UFFC_2018} or
\ndim{1} correlation methods~\cite{Denarie_TMI_2013},
and compensate for it before compounding all acquisitions
to produce a motion-compensated high-quality image.
However,
these motion compensation techniques can also suffer
from strong diffraction artifacts~\cite{Denarie_TMI_2013},
as they are themselves based on displacement estimation from low-quality images,
obtained from unfocused wavefronts.
It thus remains unclear if such methods could help improve motion estimation
in regions plagued by such artifacts.

Consequently,
there exists a great need for a robust displacement estimation technique
that does not rely on multiple acquisitions to reconstruct consecutive frames.
This is of particular interest in extreme conditions,
when analyzing rapidly evolving physical phenomena
in zones with highly heterogeneous echogenicities.

In~\cite{Perdios_ARXIV_2020a},
we introduced a method for reconstructing high-quality \gls{us} images
from single unfocused acquisitions.
It consists of a backprojection-based \gls{das} operation followed
by the application of a \gls{cnn},
specifically trained to reduce the diffraction artifacts
inherent to the deployed ultrafast \gls{us} imaging setup.
Strong artifact reduction was demonstrated in simulated,
\gls{invitro},
and \gls{invivo} environments.
The \gls{cnn}-based image reconstruction method works strictly on a
frame-by-frame basis and relies on the spatial information of each image only.
Hence,
it is completely agnostic to displacements that may occur
between consecutive frames,
making it a perfect fit for combination with state-of-the-art
image-based displacement estimation techniques.
In a preliminary work~\cite{Perdios_IUS_2019}
we showed that a \gls{cnn}-based image reconstruction method
may preserve the time-coherence of speckle patterns between consecutive frames,
which is essential to any image-based displacement estimation technique.

In this work,
we propose an approach for estimating \ndim{2} inter-frame displacements
at maximum frame rates,
by combining our \gls{cnn}-based
image reconstruction method~\cite{Perdios_ARXIV_2020a}
with a state-of-the-art \ndim{2} speckle tracking algorithm.
Although estimating the axial displacement (only) remains the standard
in \gls{us} imaging,
\ndim{2} displacement estimation is increasingly gaining attention
in both flow and tissue motion applications~\cite{Jensen_UFFC_2016_VFIb,
Fadnes_UMB_2014,Voigt_EHJCI_2015},
as it enables the analysis of more complex motion patterns.
In elastography,
\ndim{2} displacement maps may be of interest to increase the quality and
robustness of the estimated elasticity maps~\cite{Tanter_UFFC_2002}.
Also,
\ndim{2} speckle tracking represents an optimal fit for high-frame-rate
displacement estimation since,
unlike vector Doppler techniques,
it does not rely on multi-angle acquisitions.
Moreover,
displacement estimation can be performed accurately
from two consecutive frames only,
whereas Doppler-based techniques usually require multiple consecutive frames
to estimate the phase accurately.

Since our aim is to tackle displacement estimation at maximum frame rates,
the proposed approach relies only on single unfocused acquisitions
to reconstruct consecutive frames and on two consecutive frames only
to obtain \ndim{2} displacement estimates.
The primary goal of this work is to assess whether
the diffraction artifact reduction and speckle restoration capabilities
of our \gls{cnn}-based image reconstruction method~\cite{Perdios_ARXIV_2020a}
can enable accurate estimation of displacements in zones
initially shadowed by \gls{gl}, \gls{sl}, and \gls{ew} artifacts.
This work was conducted in the context of \gls{pw} imaging
with a linear transducer array (\cref{sec:methods}).
The accuracy of the proposed approach was evaluated both in
numerical and in \gls{invivo} experiments,
and was compared with a state-of-the-art \gls{cpwc}-based
displacement estimation approach (\cref{sec:experiments}).
The results obtained (\cref{sec:results}) demonstrate that the proposed approach
is capable of estimating displacements in zones initially shadowed by \gls{sl}
and \gls{gl} artifacts accurately.
However,
only slight improvements were observed in zones initially shadowed
by \gls{ew} artifacts,
which still prevent accurate displacement estimates.
In-depth results, implications, and limitations of the experiments carried out
are analyzed and discussed in \cref{sec:results,sec:discussion},
respectively.
Concluding remarks are given in \cref{sec:conclusion}.

\section{Materials and Methods}%
\label{sec:methods}

\subsection{Imaging Configurations}%
\label{sec:methods:imaging-configurations}

We considered a \gls{us} acquisition system composed of
a \gls{ge9ld} transducer
(\gls{ge-healthcare}, \gehealthcareaddress{})
and a \gls{vantage256} system
(\gls{verasonics}, \verasconicsaddress{}),
identical to the one considered in~\cite{Perdios_ARXIV_2020a}.
Relevant imaging configuration parameters are summarized in
\cref{tab:imaging-configurations}.
The \gls{ge9ld} is a \geelemnumber{}-element linear transducer array
with a center frequency of \gecenterfreq{}
and a bandwidth of \gebandwidth{} (at \SI{-6}{\decibel}),
and is commonly used for vascular imaging.
All pulse-echo acquisitions were carried by transmitting
a single-cycle tri-state waveform of \SI{67}{\percent} duty cycle
centered at \( \vsxtxfreq \),
with leading and trailing equalization pulses of quarter-cycle durations
and opposite polarities.
The received echo signals were sampled at \( \vsxrxfreq \),
guaranteeing a Nyquist sampling rate up to a bandwidth of \SI{200}{\percent}.
To reconstruct images up to a depth of \SI{60}{\milli\meter},
we considered a maximum \gls{prf} of \vsxprf{}.

\begin{table}[t]
    \sffamily
    \centering
    \caption{Specifications of the Imaging Configurations Considered}%
    \label{tab:imaging-configurations}
    \newcommand*{\tnoteelemwidthmark}{a}
\newcommand*{\tnoteeqpulsemark}{b}
\begin{threeparttable}
\begin{tabular*}{0.7\columnwidth}{l @{\extracolsep{\fill}} c}
    \toprule
    \textbf{Parameter}
    & \textbf{Value}
    \\
    \midrule
    Center frequency
    & \gecenterfreq{}
    \\
    Bandwidth
    & \gebandwidth{}
    \\
    Aperture
    & \geaperture{}
    \\
    Element number
    & \geelemnumber{}
    \\
    Pitch
    & \gepitch{}
    \\
    Element width\tnote{\tnoteelemwidthmark}
    & \geelemwidth{}
    \\
    Element height
    & \geelemheight{}
    \\
    Elevation focus
    & \geelevationfocus{}
    \\
    Transmit frequency
    & \vsxtxfreq{}
    \\
    Excitation cycles\tnote{\tnoteeqpulsemark}
    & \num{1}
    \\
    Sampling frequency
    & \vsxrxfreq{}
    \\
    \bottomrule
\end{tabular*}
\begin{tablenotes}
    \item[\tnoteelemwidthmark]
        Guessed (no official data available).
    \item[\tnoteeqpulsemark]
        Single excitation cycle with equalization pulses.
\end{tablenotes}
\end{threeparttable}%

\end{table}

All image reconstruction methods considered in this study rely on
\gls{pw} acquisitions performed without transmit apodization.
Single \gls{pw} acquisitions with normal incidence were used
for the proposed \gls{cnn}-based image reconstruction method
(\cref{sec:methods:image-reconstruction}),
and steered \gls{pw} acquisitions were used for \gls{cpwc}-based
comparison methods (\cref{sec:methods:comparison-methods}).
For each transmit-receive event,
echo signals were recorded on all transducer elements
(\gls{ie} full aperture).
A typical speed of sound in soft tissue
of \SI{1540}{\meter\per\second} was assumed,
resulting in an element spacing (\gls{ie} pitch)
of \( \num{\sim{} 0.78} \wavelength \) at the transmit frequency.
As a result,
images reconstructed with this transducer in the context of ultrafast imaging
by conventional \gls{das} algorithms will inevitably be contaminated
by \gls{gl} artifacts.
As discussed in~\cite{JensenJonas_UFFC_2016},
most linear transducer arrays available commercially were optimized for
line-by-line scanning,
and are thus suboptimal when used in the context of ultrafast imaging.
Nonetheless,
these transducer arrays remain commonly used in ultrafast
imaging~\cite{Montaldo_UFFC_2009,Tanter_UFFC_2014,JensenJonas_UFFC_2016},
thanks to their wide aperture and resulting high lateral resolution.

\subsection{CNN-Based Image Reconstruction Method}%
\label{sec:methods:image-reconstruction}

To obtain high-quality images from single unfocused acquisitions,
we relied on our \gls{cnn}-based image reconstruction method
proposed in~\cite{Perdios_ARXIV_2020a},
briefly summarized hereafter.

The method consists of first reconstructing a (vectorized) low-quality estimate
\( \imvecapprox \in \imvecdom \)
from the (vectorized) transducer elements measurements
\( \rdvecdef \),
obtained from a single unfocused insonification,
by means of a backprojection-based \gls{das} operator
\( \dasopmatdef \) as \( \imvecapprox = \dasopmat \rdvec \).
The operator \( \dasopmat \) is composed of the adjoint
of a linear measurement model (backprojection)
and a pixel-wise reweighing operator (image equalization).
The measurement model is based on linear acoustics
and is derived from the \gls{sir} model~\cite{Jensen_JASA_1991},
assuming far-field approximation both for the transmitter
(\gls{eg} ideal wavefront)
and the receiver (\gls{eg} narrow transducer element),
an ideal Dirac pulse-echo waveform,
and neglecting tissue attenuation.
Before summation,
measurement values were interpolated using a \gls{bspline} approximation
of degree three~\cite{Thevenaz_TMI_2000}.
Analytic (complex) images,
also called \gls{iq} images,
were reconstructed on a \( \imagesampling \) (Cartesian) grid,
with a width spanning the \gls{ge9ld} aperture (\cref{tab:imaging-configurations})
and a depth from \SIrange{1}{60}{\milli\meter}.
The image grid resolution was chosen to guarantee Nyquist sampling
of \gls{rf} content of \gls{us} images in both dimensions,
resulting in images of \( \numimdim \) pixels.
The process was implemented with \gls{pyus},\footnote{\pyusurl}
a \gls{gpu}-accelerated \gls{python} package for \gls{us} imaging
developed in our laboratory.

In a second step,
the low-quality estimate \( \imvecapprox \)
is fed to a \gls{cnn} \( \cnndef \),
with parameters \( \cnnparams \),
trained to recover a high-quality estimate as
\( \estimate{\imvec} = \cnn(\imvecapprox) \),
with strongly reduced diffraction artifacts and well-preserved speckle patterns.
The \gls{cnn} architecture is based on
the popular \gls{unet}~\cite{Ronneberger_MICCAI_2015}
and on~\cite{Jin_TIP_2017},
with several improvements
such as the use of \glspl{res-conv-block}
and additive intrinsic skip connections~\cite{Perdios_ARXIV_2020a}.
It is a residual \gls{cnn} with multi-scale
and multi-channel filtering properties,
composed of \ndim{2} \glspl{conv-layer} and \glspl{relu}
arranged in symmetric downsampling and upsampling paths.
As real-time displacement estimation was not a primary goal of this work,
we used the best-performing \gls{cnn} architecture
analyzed in~\cite{Perdios_ARXIV_2020a},
with \num{32} initial expansion channels.
The \gls{cnn} was trained precisely as detailed in~\cite{Perdios_ARXIV_2020a},
namely in a supervised manner using a dataset composed of
\num{30000} simulated image pairs (\gls{ie} input and ground-truth).
The well-known Adam optimizer~\cite{Kingma_ARXIV_2014} was used
to minimize the \gls{mslae} loss,
introduced in~\cite{Perdios_ARXIV_2020a} to account for both the \gls{hdr}
and the \gls{rf} property of \gls{us} images.
A total of \num{500000} iterations were performed with a batch size of two
and a learning rate of \num{5e-5}.
The same training dataset of simulated images was used.
It is composed of low-quality input images reconstructed from single \gls{pw}
acquisitions with normal incidence.
High-quality reference images were reconstructed from the complete set of
\gls{sa} acquisitions using a spatially oversampled version of the
transducer array to ensure the absence of \gls{gl} artifacts
(only possible in a simulation environment).
To reconstruct both input and reference images,
element raw-data were simulated using an in-house
\ndim{3} \gls{sir} simulator,
validated against the well-known \gls{fieldii} simulator~\cite{Jensen_MBEC_1996}.
Each numerical phantom was composed of random scatterers
with a density that ensured fully developed speckle patterns throughout the
resulting images.
The simulated images composing the training dataset are characterized by
overlapping ellipsoidal zones of random size, position, and orientation,
with mean echogenicities spanning an \num{80}-\si{\decibel} range.

\subsection{Comparative Image Reconstruction Methods}%
\label{sec:methods:comparison-methods}

For the \gls{cpwc}-based comparison methods,
acquisitions to reconstruct consecutive frames consisted of sequential
transmit-receive events of \( \cpwcanglenumber \)
differently steered \glspl{pw},
fired at maximum \gls{prf}.
The \gls{pw} steering angle spacing
was evaluated as~\cite{Montaldo_UFFC_2009,Denarie_TMI_2013}
\begin{align}
    \cpwcanglespacing
    =
    \arcsin\parens[\Big]{\frac{\wavelength}{\aperture}}
    \approx
    \SI{0.38}{\degree}
    ,
     \label{eq:cpwc-angle-spacing}
\end{align}
where \(\wavelength \) is the wavelength of transmit excitation
and \(\aperture \) is the transducer aperture.
We restricted ourselves to odd acquisition numbers,
thus the linearly increasing sequence of steering angles can be expressed as
\begin{align}
    \angleind
    =
    n \cpwcanglespacing,
    \quad
    n
    =
    -\cpwcanglemaxindex, -\cpwcanglemaxindex + 1,
    \dots, 0, \dots,
    \cpwcanglemaxindex - 1, \cpwcanglemaxindex
    \label{eq:angle-index}
    ,
\end{align}
where \(\cpwcanglemaxindex = (\cpwcanglenumber - 1) / 2 \).
We deployed an alternate steering angle sequence
\(
    (
        -\cpwcangle_{\cpwcanglemaxindex},\allowbreak
        \cpwcangle_{\cpwcanglemaxindex},\allowbreak
        -\cpwcangle_{\cpwcanglemaxindex - 1},\allowbreak
        \cpwcangle_{\cpwcanglemaxindex - 1},\allowbreak
        \dots,\allowbreak
        -\cpwcangle_{1},\allowbreak
        \cpwcangle_{1},\allowbreak
        0
    )
\),
as proposed in~\cite{Denarie_TMI_2013}.

\begin{figure}[t]
    \centering
    \includegraphics{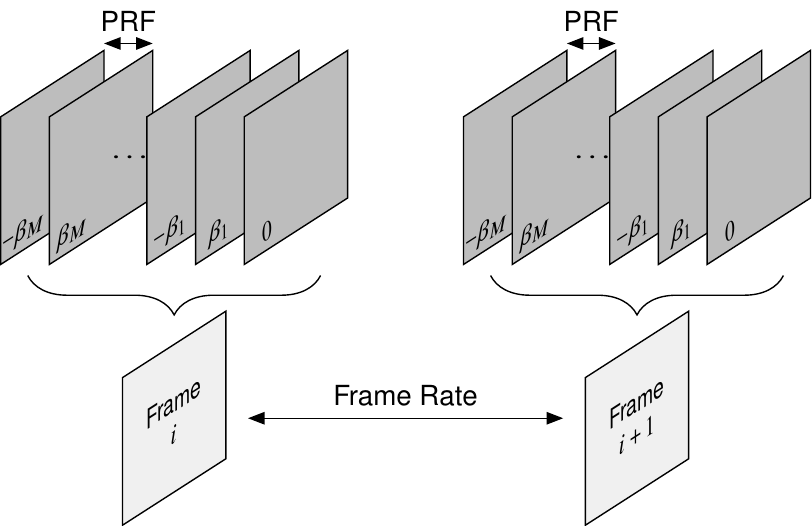}
    \caption{
        Sketch of the acquisition schemes deployed for the different
        \glsxtrfull{pw} imaging configurations considered.
        To form a single frame,
        a sequence of echo-signals from differently steered \glspl{pw}
        is acquired at a \glsxtrfull{prf} of \vsxprf{}.
        The number of \glspl{pw} composing each acquisition sequence depends
        on the imaging configuration (\cref{tab:imaging-acquisition-sequences}).
        The resulting frame rate is determined by the time interval
        between consecutive acquisition sequences,
        and is limited by the duration of a single acquisition sequence.
    }%
    \label{fig:methods:acquisition-sketch}
\end{figure}

In particular,
we considered single-\gls{pw} acquisitions with normal incidence,
used both with the proposed \gls{cnn}-based image reconstruction method
and with \gls{das} beamforming,
as well as sequences of \numlist{3;9;15;87} steered \gls{pw} acquisitions
used with \gls{das} beamforming.
Comparison \gls{das}-based methods are denoted \cpwcmethodname{1},
\cpwcmethodname{3}, \cpwcmethodname{9}, \cpwcmethodname{15},
and \cpwcmethodname{87}.
The parameters for each imaging acquisition sequence considered are
summarized in \cref{tab:imaging-acquisition-sequences};
the corresponding maximum achievable frame rates,
given the deployed \gls{prf} of \vsxprf{},
are also provided.
A sketch of the imaging acquisition schemes
is depicted in \cref{fig:methods:acquisition-sketch}.

The \cpwcmethodname{87} was used for reference purposes only and
exclusively in settings where motion artifacts were negligible.
This reference number of acquisitions was computed
following~\cite{Montaldo_UFFC_2009} as
\begin{align}
    \cpwcanglenumberref
    =
    \frac{\aperture}{\wavelength \fnumber}
    \approx
    87
    \label{eq:cpwc-reference-angle-number}
    ,
\end{align}
with an F-number \( \fnumber = 1.75 \).
The other comparison methods,
namely \cpwcmethodname{1} to \cpwcmethodname{15},
were selected to obtain a range of maximum achievable frame rates,
namely from \SIrange[]{9}{0.6}{\kilo\hertz},
spanning typical values necessary for analyzing rapid events
occurring in the human body.

\begin{table}[t]
    \sffamily
    \centering
    \caption{Plane Wave Imaging Acquisition Sequences Considered}%
    \label{tab:imaging-acquisition-sequences}
    \newcommand*{\tnotesinglepwmark}{a}
\begin{threeparttable}
\begin{tabular}{l c c c c c c c}
    \toprule
    \multirow{2}{*}{\textbf{Method}}
    & \multicolumn{5}{c}{\textbf{Sequence Parameters}}
    & \textbf{Maximum}
    \\
    & \( \cpwcanglenumber \)
    & \( \cpwcanglespacing \)
    & \( \cpwcangle_{\cpwcanglemaxindex} \)
    & Type
    & \acrshort{prf}
    & \textbf{Frame Rate}
    \\
    \midrule
    \glsxtrshort{cnn}
    & \num{1}
    & \invalidtabentry{}\tnote{\tnotesinglepwmark}
    & \invalidtabentry{}\tnote{\tnotesinglepwmark}
    & \invalidtabentry{}\tnote{\tnotesinglepwmark}
    & \invalidtabentry{}\tnote{\tnotesinglepwmark}
    & \SI{9}{\kilo\hertz}
    \\
    \cpwcmethodname{1}
    & \num{1}
    & \invalidtabentry{}\tnote{\tnotesinglepwmark}
    & \invalidtabentry{}\tnote{\tnotesinglepwmark}
    & \invalidtabentry{}\tnote{\tnotesinglepwmark}
    & \invalidtabentry{}\tnote{\tnotesinglepwmark}
    & \SI{9}{\kilo\hertz}
    \\
    \cpwcmethodname{3}
    & \num{3}
    & \SI{0.38}{\degree}
    & \SI{0.38}{\degree}
    & Alternate
    & \SI{9}{\kilo\hertz}
    & \SI{3}{\kilo\hertz}
    \\
    \cpwcmethodname{9}
    & \num{9}
    & \SI{0.38}{\degree}
    & \SI{1.52}{\degree}
    & Alternate
    & \SI{9}{\kilo\hertz}
    & \SI{1}{\kilo\hertz}
    \\
    \cpwcmethodname{15}
    & \num{15}
    & \SI{0.38}{\degree}
    & \SI{2.66}{\degree}
    & Alternate
    & \SI{9}{\kilo\hertz}
    & \SI{0.6}{\kilo\hertz}
    \\
    \cpwcmethodname{87}
    & \num{87}
    & \SI{0.38}{\degree}
    & \SI{16.34}{\degree}
    & Alternate
    & \SI{9}{\kilo\hertz}
    & \SI{0.1}{\kilo\hertz}
    \\
    \bottomrule
\end{tabular}
\begin{tablenotes}
    \item[\tnotesinglepwmark] Single \gls{pw} with normal incidence.
\end{tablenotes}
\end{threeparttable}%

\end{table}

Each \gls{pw} acquisition was reconstructed using the \gls{das} algorithm
detailed in \cref{sec:methods:image-reconstruction}.
Coherent compounding of images reconstructed from steered acquisitions
was realized by simple pixel-wise averaging.
Note that as \cpwcmethodname{1} only relies on single-\gls{pw} acquisitions,
it is not a compounding method.
Its designation was adopted to simplify the naming convention.
Also,
images obtained from \cpwcmethodname{1} are identical to input images
of the \gls{cnn}-based image reconstruction
(\cref{sec:methods:image-reconstruction}),
as the same \gls{das} algorithm was deployed in both cases.

\subsection{Speckle Tracking Algorithm}%
\label{sec:methods:speckle-tracking}

\glsunset{ieee}

The proposed speckle tracking algorithm is
a block-matching algorithm based on normalized cross-correlation.
It is heavily inspired by both the speckle tracking method described
in~\cite{Perrot_IUS_2018},
which won the challenge on \gls{sa-vfi}
organized during the 2018 \gls{ieee} \gls{ius}~\cite{Jensen_IUS_2018},
and the \gls{pivlab} toolbox~\cite{Thielicke_JORS_2014},
a popular software for \gls{piv}.
Speckle tracking is fundamentally linked to \gls{piv}.
However,
instead of tracking particles to visualize flows,
speckle tracking estimates displacements by tracking speckle patterns
arising from interferences by scatterers separated by sub-resolution distances,
assuming that these patterns are highly correlated between consecutive frames.

To estimate the \ndim{2} displacement field between two consecutive frames
\( \pivframe_{1} \) and \( \pivframe_{2} \),
both frames were identically subdivided into overlapping interrogation
windows.
The most probable displacement that occurred between a pair of interrogation
windows was obtained by finding the maximum value (peak)
of the (\ndim{2}) \gls{zncc}.
To achieve sub-pixel precision,
we applied a \ndim{2} Gaussian regression around the \gls{zncc} peak,
as proposed in~\cite{Nobach_EF_2005}.
In order to analyze complex displacements,
including shear and rotation,
this process was deployed in a coarse-to-fine multi-pass
algorithm~\cite{Thielicke_JORS_2014}.
Between each pass,
\( \pivframe_{2} \) was deformed (\gls{bspline} interpolation)
using the estimated displacements to resemble \( \pivframe_{1} \) more closely.
For the next pass,
the displacements between \( \pivframe_{1} \)
and the deformed \( \pivframe_{2} \) were estimated in a similar way.
The remaining displacement estimates of each pass were accumulated,
resulting in more accurate estimates after a few passes.
After each pass,
statistical outliers of the estimates
were smoothed using the unsupervised smoothing algorithm
described in~\cite{Garcia_CSDA_2010}.

Speckle tracking was performed on envelope images,
obtained by computing the (pixel-wise) modulus of \gls{iq} images.
Envelope images were downsampled by a factor of two in the axial dimension,
in a uniformly spaced spatial grid of
\( \envsampling \)
(\gls{ie} \( \numenvdim \) pixels).
While applying normalized cross-correlation-based speckle tracking directly to
\gls{rf} signals may lead to a higher precision than using envelope
signals~\cite{Walker_UFFC_1994},
especially when analyzing very small displacements close to the Cramér-Rao
lower bound~\cite{Walker_UFFC_1995},
it is also much more prone to faulty displacement estimation because of speckle
decorrelation~\cite[Sec.~14.2.1]{Loizou_BOOK_2018}.
Speckle decorrelation increases when analyzing larger displacements,
more complex displacements patterns with strong gradients (\gls{eg} rotation),
and tissue deformation~\cite{Bohs_ULTRAS_2000,Meunier_TMI_1995}.
As our method is designed to be a robust displacement estimator over
a wide range of displacements and flow patterns,
envelope images were preferred for the purpose of speckle tracking.
However,
it is easily adapted to work with \gls{rf} images if the potential increase
in precision for small displacements is of interest.

For adapting the speckle tracking parameters to the imaging configurations
and displacement ranges considered,
we cross-validated a wide range of different
interrogation window sizes, number of passes, and window overlaps
using a dedicated numerical test phantom,
namely a rotating cylinder centered at the elevation focus of the transducer,
similar to the ones deployed in the numerical experiment
(\cref{sec:experiments:numerical-experiment}).
Two different angular velocities were considered,
resulting in the same inter-frame displacements considered in this work.
Consecutive frames were generated by simulating high-quality images
using \cpwcmethodname{87} without rotating the cylinder between
successive steered \gls{pw} acquisitions
(only achievable in a simulation environment).
This strategy of \enquote{pausing} motion during a complete
compounded acquisition sequence was exclusively deployed for the purpose
of finding optimal speckle tracking parameters,
to avoid being biased by potential motion artifacts.
Inter-acquisition motion was considered in the following numerical experiment
(\cref{sec:experiments:numerical-experiment}).

Interestingly,
the speckle tracking parameters yielding best overall displacement estimates
in our settings were identical to the ones deployed in~\cite{Perrot_IUS_2018}.
Thus,
for all experiments conducted in this work,
irrespectively of the displacement range and frame rate under consideration,
we deployed the proposed speckle algorithm with four passes,
square interrogation windows of \SIlist{4;2.5;2;1.5}{\milli\meter},
and a window overlap of \SI{65}{\percent}.

\subsection{Metrics}%
\label{sec:methods:metrics}

To evaluate the accuracy of displacement estimates throughout the experiments,
we relied on the \gls{repe},
a normalized version of the well-known \gls{epe},
commonly used in flow estimation techniques~\cite{Otte_ECCV_94,Baker_IJCV_2011}.
Considering a displacement estimate vector
\( \estimate{\dispvector} \in \set{R}^2 \)
and its true counterpart \( \dispvector \in \set{R}^2 \),
the \gls{repe} can be expressed as
\begin{align}
    \text{\glsxtrshort{repe}}
    =
    \frac{
        \twonorm{\estimate{\dispvector} - \dispvector}
    }{
        \twonorm{\dispvector}
    }
    \label{eq:repe}
    ,
\end{align}
where \( \twonorm{\argdot} \) represents the Euclidean norm.
The main advantage of \gls{repe} over \gls{epe} comes from its relative nature,
enabling a more reliable comparison over a wide range of displacements.
On the other hand,
\gls{repe} becomes unstable as the reference displacement tends to zero.
Such cases should therefore be analyzed with care.

We also relied on the \gls{mrepe} as a global metric to assess a set of
\( \dsetsize \) displacement estimates and true counterparts
\(
    \setroster{
        \braces{
            \estimate{\dispvector}_{1}
            ,
            \dispvector_{1}
        },
        \ellipsis{},
        \braces{
            \estimate{\dispvector}_{\dsetsize}
            ,
            \dispvector_{\dsetsize}
        }
    }
\)
(\gls{eg} extracted from a region of interest),
by simply computing the sample mean of all \gls{repe} values over the set.

\section{Experiments}%
\label{sec:experiments}

We conducted two experiments (numerical and \gls{invivo}) to assess
the performance of the proposed \ndim{2} displacement estimation approach,
which combines
our \gls{cnn}-based image reconstruction method~\cite{Perdios_ARXIV_2020a}
(\cref{sec:methods:image-reconstruction})
to reconstruct consecutive frames from single-\gls{pw} acquisitions
and the deployed speckle tracking algorithm
(\cref{sec:methods:speckle-tracking}).
In both experiments,
we compared the proposed \gls{cnn}-based displacement estimation method to
\gls{cpwc}-based tracking,
which consists of applying the same speckle tracking algorithm
to consecutive frames reconstructed using conventional \gls{cpwc}
(\cref{sec:methods:comparison-methods}).
For \gls{cpwc},
a larger number of compounded acquisitions results,
if motion artifacts are negligible,
in better image quality and consequently in improved displacement estimation,
at the cost of a reduced achievable frame rate.
Thus,
by studying different numbers of compounded acquisitions
(\cref{tab:imaging-acquisition-sequences})
we compared the proposed approach to multiple levels of displacement estimation
accuracy.

\subsection{\sectitlenamenumexp}%
\label{sec:experiments:numerical-experiment}

For the first experiment,
we used computer simulations to control the motion pattern,
the relative echogenicities of tissue-mimicking structures,
and the diffraction artifact levels precisely.
The goal is to evaluate the quality of displacement tracking
that can be achieved using the proposed method in rapidly moving,
highly heterogeneous tissue,
where strong diffraction artifacts hinder proper motion analysis
with conventional \gls{cpwc}-based tracking.
All simulations were conducted using the same \gls{sir} simulator
used to generate the training dataset (\cref{sec:methods:image-reconstruction}).

\begin{figure*}[t]
    \centering
    \includegraphics{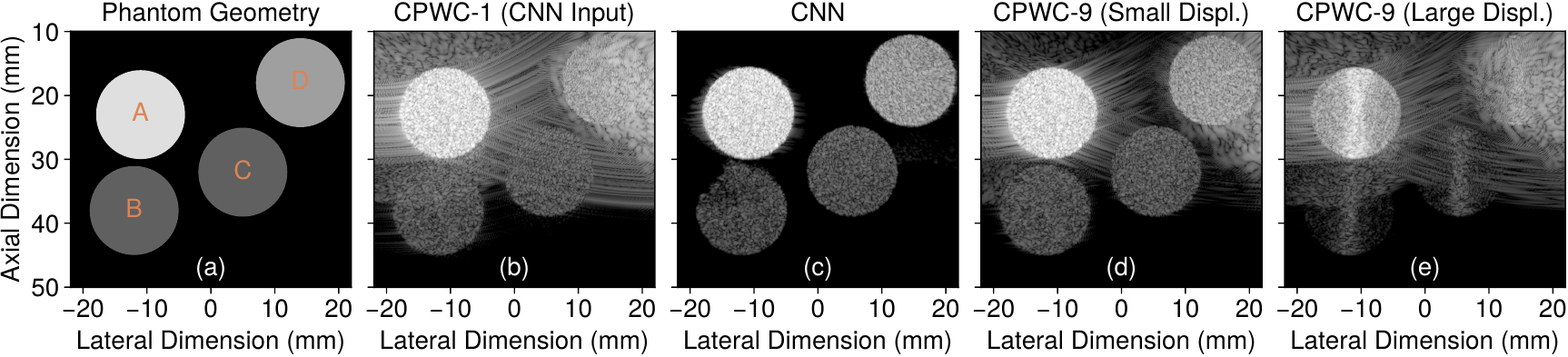}%
    {\phantomsubcaption\label{fig:experiments:numerical-phantom:mask}}%
    {\phantomsubcaption\label{fig:experiments:numerical-phantom:pw}}%
    {\phantomsubcaption\label{fig:experiments:numerical-phantom:cnn}}%
    {\phantomsubcaption\label{fig:experiments:numerical-phantom:cpwc-9-sd}}%
    {\phantomsubcaption\label{fig:experiments:numerical-phantom:cpwc-9-ld}}%
    \caption{%
        \Gls{bmode} image representations
        (\num{80}-\si{\decibel} range)
        of a numerical test phantom sample:
        \subref{fig:experiments:numerical-phantom:mask}
        the \ndim{2} geometry of the deployed numerical phantoms,
        composed of four cylinders (A, B, C, and D)
        filled with dense point-scatterers rotating
        at constant angular velocity around their respective cylinder center;
        \subref{fig:experiments:numerical-phantom:pw}
        image reconstructed by \glsxtrfull{das} beamforming a single
        plane-wave (\glsxtrshort{pw}) acquisition (\cpwcmethodname{1}),
        simultaneously representing the \glsxtrfull{cnn} input image
        for the proposed method;
        \subref{fig:experiments:numerical-phantom:cnn}
        image reconstructed using \gls{cnn}-based reconstruction;
        images reconstructed by \glsxtrfull{cpwc} using nine
        steered \gls{pw} acquisitions (\cpwcmethodname{9}):
        \subref{fig:experiments:numerical-phantom:cpwc-9-sd}
        small displacement range
        and
        \subref{fig:experiments:numerical-phantom:cpwc-9-ld}
        large displacement range.
        The frame rate and displacement range for each
        image reconstruction method considered are given
        in \cref{tab:numerical-phantom-rotation-settings}.
    }%
    \label{fig:experiments:numerical-phantom}
\end{figure*}

We designed a dynamic numerical test phantom composed of scatterers
randomly positioned within four cylinders
[\( \edtnumphtcylAlbl \), \( \edtnumphtcylBlbl \), \( \edtnumphtcylClbl \),
and \( \edtnumphtcylDlbl \) in \cref{fig:experiments:numerical-phantom:mask}],
embedded in an anechoic background.
Each cylinder has a radius of \zylradius{}
and a height of \zylheight{},
the latter corresponding to the resolution cell size in elevation
evaluated for the imaging configuration considered~\cite{Perdios_ARXIV_2020a}.
Within each of the four zones,
an average of \sprctext{} scatterers per resolution cell was used
to ensure fully developed speckle patterns
in the resulting images~\cite[Sec.~8.4.4]{Szabo_BOOK_2014}.
The cylinders were centered such that cylinder \( \edtnumphtcylAlbl \) spawns
distinct and spatially separable diffraction artifacts onto
cylinders \( \edtnumphtcylBlbl \), \( \edtnumphtcylClbl \),
and \( \edtnumphtcylDlbl \).
Cylinders \( \edtnumphtcylBlbl \), \( \edtnumphtcylClbl \),
and \( \edtnumphtcylDlbl \) were positioned such that they are maximally covered
by \gls{ew}, \gls{sl}, and \gls{gl} artifacts, respectively
[\cref{fig:experiments:numerical-phantom:pw}].
The mean amplitudes of scatterers located within cylinders
\( \edtnumphtcylBlbl \), \( \edtnumphtcylClbl \), and \( \edtnumphtcylDlbl \)
were chosen to blend in with the amplitude of \gls{ew}, \gls{sl}, and \gls{gl}
artifacts arising from cylinder \( \edtnumphtcylAlbl \)
[\cref{fig:experiments:numerical-phantom:pw}].
Specifically,
the mean amplitudes in cylinders \( \edtnumphtcylAlbl \),
\( \edtnumphtcylBlbl \), \( \edtnumphtcylClbl \), and \( \edtnumphtcylDlbl \)
were set to \zylechoa{}, \zylechob{}, \zylechoc{}, and \zylechod{}
with respect to an arbitrary \SI{0}{\decibel} reference,
respectively.
Between successive simulated transmit-receive events
(\gls{ie} steered \glspl{pw}),
the scatterers were rotated
with a constant counter-clockwise angular velocity
around the center of the cylinder within which they are positioned.
The same angular velocity was used for all cylinders.

This experiment is designed to evaluate the accuracy of displacement estimates,
obtained using the same speckle tracking algorithm on consecutive frames
reconstructed with the different image reconstruction methods considered.
For this purpose,
displacements were estimated using the proposed \gls{cnn}-based approach,
as well as \cpwcmethodname{1}, \cpwcmethodname{3}, \cpwcmethodname{9},
and \cpwcmethodname{15}.
Each method was deployed at its maximum achievable frame rate
(\cref{tab:imaging-acquisition-sequences}),
while always considering the same range of inter-frame displacements
for comparison purposes.
Inter-frame displacements ranging from \disprangetot{}
(\gls{ie} approximately from \( \wavelength / 10 \) to \( 2 \wavelength \))
were analyzed,
covering a range from the small displacements that typically occur
in shear-wave elastography~\cite{Montaldo_UFFC_2009}
or acoustic radiation force imaging~\cite{Pinton_UFFC_2006},
up to the large displacements that typically occur in external
compression-based elastography~\cite{Pinton_UFFC_2006}.
It can be noted that these inter-frame displacement ranges correspond to
velocities up to \SI{5.4}{\meter\per\second} for the two methods capable of
achieving a frame rate of \SI{9}{\kilo\hertz} in these settings
(\gls{ie} \cpwcmethodname{1} and \gls{cnn}).
Such velocities are close to peak velocities inside the cardiovascular
system~\cite{Routh_EMB_1996}.

Two different sets of numerical phantoms were simulated
for each image reconstruction method considered and associated frame rate,
covering two inter-frame displacement ranges,
namely \disprangesmall{} (small displacement range)
and \disprangelarge{} (large displacement range).
The respective angular velocities were determined such that the
maximum inter-frame displacement occurs at a radius of \zylradcropp{}.
The remaining border of \zylborder{}
was ignored to avoid speckle tracking border effects in the quality evaluation.
It corresponds to the approximate average resolution cell size
in the transducer plane.
A similar zone was ignored in the center of each cylinder.
Displacement ranges are made explicit in
\cref{tab:numerical-phantom-rotation-settings}
for each image reconstruction method considered,
and the corresponding cross-radial velocity ranges are also provided
as additional information.
It can be noted that the large-displacement case involves displacements greater
than half the deployed wavelength.
As a result,
motion artifacts are expected for \gls{cpwc} methods~\cite{Denarie_TMI_2013}.

\begin{table}[t]
    \sffamily
    \centering
    \caption{%
        Displacement and Velocity Ranges Considered\\
        for the Numerical Experiment
    }%
    \label{tab:numerical-phantom-rotation-settings}
    \newcommand*{\disprangelargetab}{\numrange[range-phrase = --]{33}{600}}
\newcommand*{\disprangesmalltab}{\numrange[range-phrase = --]{3.3}{60}}

\begin{threeparttable}
\begin{tabular}{l c c c c c}
    \toprule
    \multirow{2}{*}{\textbf{Method}}
    & \textbf{Frame}
    & \multicolumn{2}{c}{\textbf{Large Ranges}}
    & \multicolumn{2}{c}{\textbf{Small Ranges}}
    \\
    & \textbf{Rate}
    & D.\@ (\si{\micro\meter})
    & V.\@ (\si{\centi\meter\per\second})
    & D.\@ (\si{\micro\meter})
    & V.\@ (\si{\centi\meter\per\second})
    \\
    \midrule
    \glsxtrshort{cnn}
    & \SI{9}{\kilo\hertz}
    & \disprangelargetab{}
    & \numrange[range-phrase = --]{29.7}{540}
    & \disprangesmalltab{}
    & 2.97--54\hphantom{.}
    \\
    \cpwcmethodname{1}
    & \SI{9}{\kilo\hertz}
    & \disprangelargetab{}
    & \numrange[range-phrase = --]{29.7}{540}
    & \disprangesmalltab{}
    & 2.97--54\hphantom{.}
    \\
    \cpwcmethodname{3}
    & \SI{3}{\kilo\hertz}
    & \disprangelargetab{}
    & \hphantom{0}9.9--180
    & \disprangesmalltab{}
    & 0.99--18\hphantom{.}
    \\
    \cpwcmethodname{9}
    & \SI{1}{\kilo\hertz}
    & \disprangelargetab{}
    & \numrange[range-phrase = --]{3.3}{60}
    & \disprangesmalltab{}
    & 0.33--6\hphantom{0.}
    \\
    \cpwcmethodname{15}
    & \SI{0.6}{\kilo\hertz}
    & \disprangelargetab{}
    & \numrange[range-phrase = --]{2.0}{36}
    & \disprangesmalltab{}
    & \numrange[range-phrase = --]{0.20}{3.6}
    \\
    \bottomrule
\end{tabular}
\end{threeparttable}%

\end{table}

For all test configurations considered (\gls{ie} method and displacement range),
\numframessim{} statistically independent scatterer realizations were simulated,
resulting in \numframessim{} inter-frame displacement estimate maps
for each configuration.
The accuracy of each method was measured locally in terms of \gls{repe},
by computing \cref{eq:repe} for each displacement estimate (grid point)
and corresponding true (analytical) value.
The average local \gls{repe} was also computed over the \numframessim{}
independent realizations (in each displacement estimate grid point).

\subsection{\sectitlenamephysexp}%
\label{sec:experiments:invivo-experiment}

For the second experiment,
we applied the proposed approach to \gls{invivo} acquisitions,
to analyze the natural tissue motion around the carotid artery.
The goal of this experiment is to evaluate the robustness and translatability
of the results obtained in the numerical experiment
to the full complexity of \gls{invivo} imaging.
As the natural tissue motion induced by cardiac pulsations in the vicinity of
the carotid artery is small compared with the one considered in the numerical
experiment,
similar inter-frame displacements could be studied at a much lower frame rate,
enabling the use of \cpwcmethodname{87} for obtaining
high-quality reference displacement estimates.

We analyzed the slow-moving tissue between the skin and the carotid artery
of a healthy volunteer.
In particular,
motion within a specific tissue region of size \( \exproisize \)
(\cref{fig:experiments:invivo-results})
was analyzed from consecutive frames acquired at a frame rate of \expfreq{}.
This resulted in inter-frame displacements similar to those studied in
the numerical experiment (\cref{sec:experiments:numerical-experiment}),
namely ranging from \SIrange{5}{125}{\micro\meter} approximately
[\cref{fig:experiments:invivo-results:mean-disp-cpwc-87}].
Therefore,
identical speckle tracking settings were used
(\cref{sec:methods:speckle-tracking}).
Speckle tracking was performed on full images,
but we restricted our analysis to a specific zone
characterized by fully developed speckle patterns,
plagued by diffraction artifacts mainly originating
from the highly echogenic carotid walls
when imaged using \cpwcmethodname{1}
[\cref{fig:experiments:invivo-results:pw}].
The mean echogenicity of the analyzed speckle zone was approximately
\SI{20}{\decibel} lower than the echogenicity of the carotid walls,
thus similar to the relative echogenicity between
cylinders \edtnumphtcylAlbl{} and \edtnumphtcylDlbl{} studied in the numerical
experiment.

We compared displacement estimates obtained using the proposed
\gls{cnn}-based approach, \cpwcmethodname{1} (\gls{ie} the \gls{cnn} input),
and \cpwcmethodname{15} with respect to reference displacement estimates
obtained with \cpwcmethodname{87} (\cref{tab:imaging-acquisition-sequences}).
As compounded acquisition sequences were performed at a \gls{prf}
of \SI{9}{\kilo\hertz},
motion artifacts were negligible.
More specifically,
the maximum mean displacement estimated during a complete compounded acquisition
sequence for \cpwcmethodname{87} was approximately of \SI{12}{\micro\meter}.
This amounts to approximately \( \wavelength / 25 \)
and motion artifacts can therefore be neglected~\cite{Denarie_TMI_2013}.
For each method being compared,
consecutive frames were reconstructed using the relevant subset
of steered \gls{pw}(s) acquired for the reference
\cpwcmethodname{87} method (\cref{sec:methods:comparison-methods}).

A total of \num{30} frames were obtained at a frame rate of \SI{10}{\hertz},
resulting in \num{29} inter-frame displacement estimate maps.
For each inter-frame displacement estimate map,
the accuracy of each method was measured locally in terms of \gls{repe},
by computing \cref{eq:repe} for each displacement estimate (grid point)
and corresponding reference value (\cpwcmethodname{87}).
The quality of the displacement estimates for each pair of frames
was assessed by computing the \gls{mrepe} obtained within
the region of interest.

\section{Results}%
\label{sec:results}

\begin{figure*}[t]
    \centering%
    \includegraphics{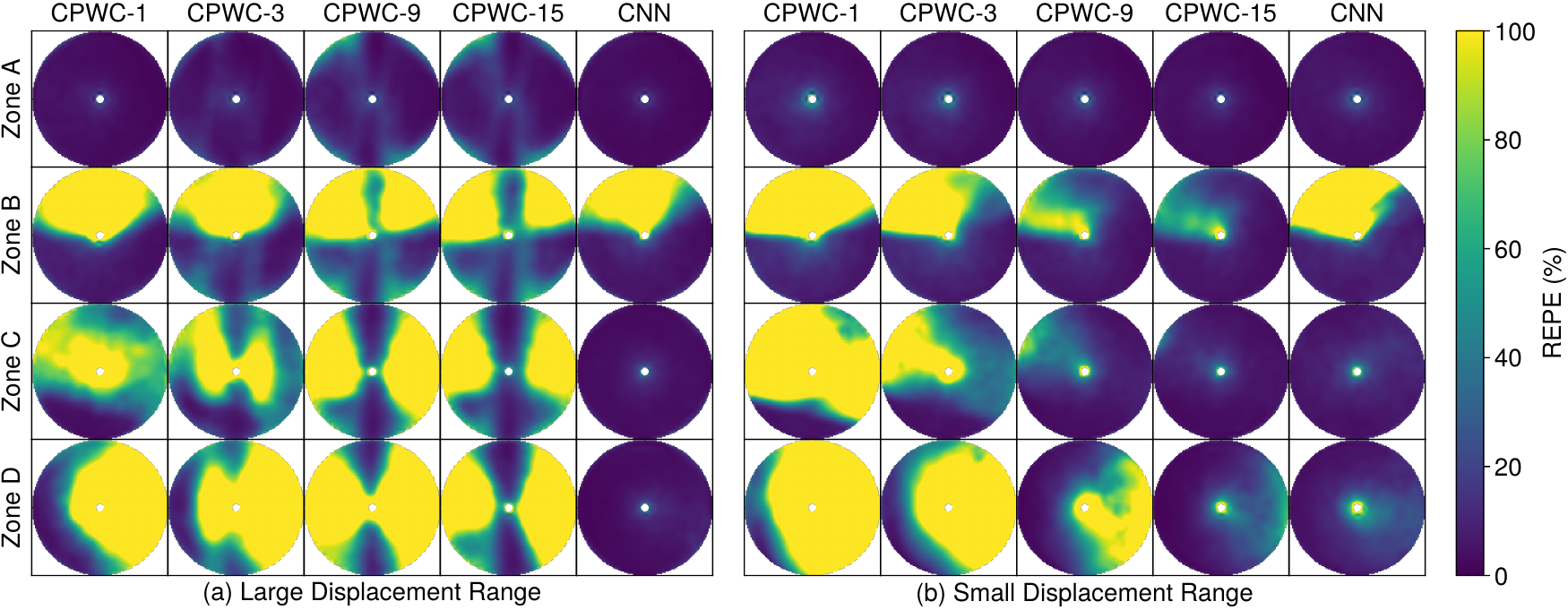}%
    {\phantomsubcaption\label{fig:results:numerical-phantom:600}}%
    {\phantomsubcaption\label{fig:results:numerical-phantom:060}}%
    \caption{%
        Local \glsxtrfull{repe},
        averaged over \num{50} independent realizations,
        of the \ndim{2} displacement estimates inside
        each of the numerical phantom zones
        [\edtnumphtcylAlbl{}, \edtnumphtcylBlbl{}, \edtnumphtcylClbl{},
        and \edtnumphtcylDlbl{}
        in \cref{fig:experiments:numerical-phantom:mask}],
        obtained by applying the deployed \ndim{2} speckle tracking algorithm
        (\cref{sec:methods:speckle-tracking})
        on two consecutive frames
        for the two inter-frame displacement ranges considered:
        \subref{fig:results:numerical-phantom:600}
        large displacement range (from \disprangelarge{});
        \subref{fig:results:numerical-phantom:060}
        small displacement range (from \disprangesmall{}).
        Consecutive frames were reconstructed either by \glsxtrfull{cpwc}
        from \numlist{1;3;9;15} differently steered
        \glspl{pw},
        or using the proposed \glsxtrfull{cnn}-based image reconstruction method
        from single \glspl{pw}.
        The frame rate and displacement range for each
        image reconstruction method considered are given
        in \cref{tab:numerical-phantom-rotation-settings}.
        The displayed \glsxtrshort{repe} range is limited to \SI{100}{\percent}.
        Local \glsxtrshort{repe} values were interpolated onto a fine grid for
        display purposes.
    }%
    \label{fig:results:loc-epe}%
\end{figure*}

\begin{table*}[t]
    \sffamily
    \centering
    \caption{%
        Global Evaluation Metrics of the Numerical Experiment
    }%
    \label{tab:numerical-phantom-global-metrics}
    \newlength{\rtabw}\setlength{\rtabw}{\widthof{\cpwcmethodname{15}}}

\newcommand{\tnotervemark}{a}
\newcommand{\tnotevmrepemark}{b}

\newcommand{\metricAlabel}{RVE (\si{\percent})}
\newcommand{\metricBlabel}{\glsxtrshort{mrepe} (\si{\percent})}
\newcommand{\metricClabel}{MEDREPE (\si{\percent})}
\newcommand{\metricDlabel}{VMREPE (\si{\percent})}
\newcommand{\metricAlabeltnote}{RVE\textsuperscript{\tnotervemark} (\si{\percent})}
\newcommand{\metricBlabeltnote}{\metricBlabel}
\newcommand{\metricClabeltnote}{\metricClabel}
\newcommand{\metricDlabeltnote}{%
VMREPE\textsuperscript{\tnotevmrepemark} (\si{\percent})%
}

\sisetup{
    round-mode = places,
    round-precision = 2,
    table-number-alignment = center,
    table-figures-integer = 3,
    table-figures-decimal = 2,
}
\begin{threeparttable}
\begin{tabular}{
        l l  S S S S S  S S S S S
    }
    \toprule
    \multirow{2}[1]{*}{\textbf{Zone}}
    & \multirow{2}[1]{*}{\textbf{Metric}}
    & \multicolumn{5}{c}{\textbf{Large Displacement Range}}
    & \multicolumn{5}{c}{\textbf{Small Displacement Range}}
    \\
    \cmidrule(lr){3-7} \cmidrule(lr){8-12}
    &
    & {\cpwcmethodname{1}}
    & {\cpwcmethodname{3}}
    & {\cpwcmethodname{9}}
    & {\cpwcmethodname{15}}
    & {\glsxtrshort{cnn}}
    & {\cpwcmethodname{1}}
    & {\cpwcmethodname{3}}
    & {\cpwcmethodname{9}}
    & {\cpwcmethodname{15}}
    & {\glsxtrshort{cnn}}
    \\
    \midrule
    \multirow{2}{*}{\edtnumphtcylAlbl{}}
    & \metricAlabel{}
    & 100.0000 & 100.0000 & 100.0000 & 100.0000 & 100.0000
    & 100.0000 & 100.0000 & 100.0000 & 100.0000 & 100.0000
    \\
    & \metricBlabeltnote{}
    &   4.4460 &   7.2406 &  12.9919 &  12.8443 &   3.6159
    &   7.3442 &   6.9130 &   5.2454 &   4.3558 &   5.8100
    \\
    \midrule
    \multirow{2}{*}{\edtnumphtcylBlbl{}}
    & \metricAlabel{}
    &  63.1038 &  69.4758 &  63.1038 &  68.8592 &  74.6146
    &  57.7595 &  73.7924 &  99.3834 &  99.6917 &  67.4203
    \\
    & \metricBlabel{}
    &  78.5797 &  61.3176 &  82.7172 &  72.6517 &  48.3588
    & 143.9140 &  64.3007 &  26.6593 &  19.3666 &  95.4429
    \\
    \midrule
    \multirow{2}{*}{\edtnumphtcylClbl{}}
    & \metricAlabel{}
    &  85.2697 &  77.8008 &  51.5560 &  65.2490 & 100.0000
    &  29.2531 &  81.6390 & 100.0000 & 100.0000 & 100.0000
    \\
    & \metricBlabel{}
    &  67.5022 &  66.2965 &  91.4927 &  71.0270 &   4.9836
    & 192.5744 &  54.3719 &  17.6430 &   8.2927 &   9.6123
    \\
    \midrule
    \multirow{2}{*}{\edtnumphtcylDlbl{}}
    & \metricAlabel{}
    &  44.5932 &  44.1813 &  34.8095 &  49.4336 & 100.0000
    &  22.1421 &  42.0185 &  82.2863 &  99.6910 &  99.5881
    \\
    & \metricBlabel{}
    & 135.1803 & 120.6853 & 123.7411 & 100.1869 &   5.5143
    & 504.3849 & 159.1545 &  52.8969 &  17.8259 &  15.6713
    \\
    \bottomrule
\end{tabular}
\end{threeparttable}%

\end{table*}

\subsection{\sectitlenamenumexp}%
\label{sec:results:numerical-experiment}

\Cref{fig:results:loc-epe} displays local \gls{repe} values,
averaged over the \num{50} independent realizations performed
in each configuration considered (\cref{sec:experiments:numerical-experiment}).
To support the analysis,
we also provide two global metrics computed for each zone, method,
and displacement range considered (\cref{tab:numerical-phantom-global-metrics}),
namely the \gls{mrepe} and the \gls{rve}.
For the \gls{rve},
a local \gls{repe} value (averaged over the \num{50} independent realizations)
exceeding \SI{100}{\percent} was deemed invalid.
It is thus directly related to the amount of saturated \gls{repe} values
depicted in \cref{fig:results:loc-epe},
and provides a global metric less sensitive than \gls{mrepe}
to potentially huge-but-scarce local \gls{repe} values.
\\\indent
Zone \edtnumphtcylAlbl{} was designed such that it did not suffer from
diffraction artifacts and could be used to assess displacement estimation
in pure speckle zones.
In the large-displacement case [\cref{fig:results:numerical-phantom:600}],
\gls{cpwc}-based tracking suffered from increasing motion artifacts
with the number of compounded acquisitions when tracking identical
inter-frame displacements (\gls{ie} at decreasing frame rates),
reaching a stable motion artifact level after nine compounded acquisitions.
The proposed method performed best and improved over
\cpwcmethodname{1} both in terms of local and global metrics.
In the small-displacement case [\cref{fig:results:numerical-phantom:060}],
motion artifacts are negligible and all methods performed efficiently.
A typical comparison of \gls{cpwc} with and without motion artifacts is
shown in \cref{fig:experiments:numerical-phantom:cpwc-9-sd,%
fig:experiments:numerical-phantom:cpwc-9-ld} for \cpwcmethodname{9}.

Zone \edtnumphtcylBlbl{} was designed to suffer from \gls{ew} artifacts.
The proposed method was not capable of restoring speckle patterns
shadowed by \gls{ew} artifacts accurately,
resulting in performance metrics only slightly
improved compared with \cpwcmethodname{1}.
Inaccurate restoration of speckle patterns plagued by \gls{ew} artifacts
can be observed in \cref{fig:experiments:numerical-phantom:cnn}
(\gls{eg} clipped values).
These artifacts could only be progressively resolved in the small displacement
case [\cref{fig:results:numerical-phantom:060}] with the increase
in compounded acquisitions,
because motion artifacts are negligible in that case.

Zone \edtnumphtcylClbl{} was designed to suffer from \gls{sl} artifacts.
In the large-displacement case [\cref{fig:results:numerical-phantom:600}],
the reduction in \gls{sl} artifacts achieved by compounding several acquisitions
was counteracted by the induced motion artifacts,
except in zones of pure lateral movement,
making proper tracking impossible using \gls{cpwc}-based tracking.
The proposed method was capable of properly estimating displacements,
with a quality only slightly worse than in artifact-free
zone \edtnumphtcylAlbl{}.
In the small-displacement case [\cref{fig:results:numerical-phantom:060}],
\gls{cpwc}-based tracking was improved with the increase
in compounded acquisitions,
thanks to a more efficient \gls{sl} reduction than with motion artifacts.
The proposed method achieved a quality slightly worse than \cpwcmethodname{15}
but significantly better than \cpwcmethodname{9}.

Zone \edtnumphtcylDlbl{} was designed to suffer from \gls{gl} artifacts,
that increase in strength towards the right edge of the image.
In the large-displacement case [\cref{fig:results:numerical-phantom:600}],
compounding multiple acquisitions reduced \gls{gl} artifacts.
Yet,
motion artifacts prevented accurate displacement estimation
except in zones of pure lateral movement.
The proposed method significantly improved the displacement estimation quality
over \cpwcmethodname{1} and was the only method to enable tracking displacements
in this case.
In the small-displacement case [\cref{fig:results:numerical-phantom:060}],
the increase in compounded acquisitions enabled \gls{cpwc}-based tracking
to reduce the effect of \glspl{gl} and restore the underlying speckle patterns,
progressively resulting in an increased \gls{rve} and lower \gls{mrepe}.
The proposed method performed slightly better than \cpwcmethodname{15}.

\subsection{\sectitlenamephysexp}%
\label{sec:results:invivo-experiment}

\begin{figure*}[t]
    \centering
    \includegraphics{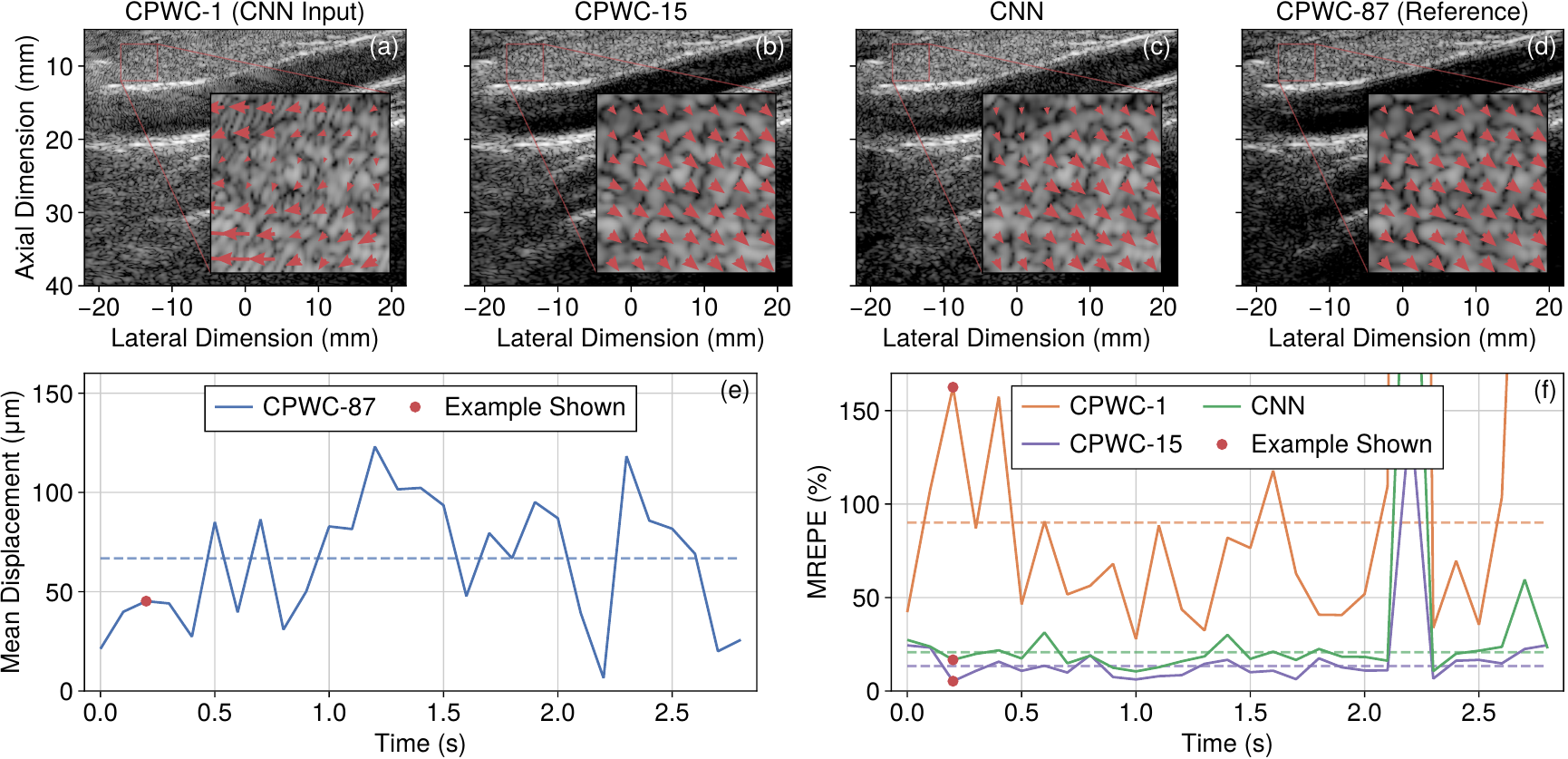}%
    {\phantomsubcaption\label{fig:experiments:invivo-results:pw}}%
    {\phantomsubcaption\label{fig:experiments:invivo-results:cpwc-15}}%
    {\phantomsubcaption\label{fig:experiments:invivo-results:cnn}}%
    {\phantomsubcaption\label{fig:experiments:invivo-results:cpwc-87}}%
    {\phantomsubcaption\label{fig:experiments:invivo-results:mean-disp-cpwc-87}}%
    {\phantomsubcaption\label{fig:experiments:invivo-results:mean-repe}}%
    \caption{%
        Examples of displacement estimates,
        mean reference displacement magnitude,
        and \glsxtrfull{mrepe},
        obtained using the displacement estimation methods considered,
        in a fully developed speckle zone
        above the carotid artery:
        images of a longitudinal view of the carotid artery,
        are shown for
        \subref{fig:experiments:invivo-results:pw}
        \cpwcmethodname{1} (also \gls{cnn} input),
        \subref{fig:experiments:invivo-results:cpwc-15}
        \cpwcmethodname{15},
        \subref{fig:experiments:invivo-results:cnn}
        \gls{cnn},
        and
        \subref{fig:experiments:invivo-results:cpwc-87}
        \cpwcmethodname{87} (reference);
        the bottom row shows
        \subref{fig:experiments:invivo-results:mean-disp-cpwc-87}
        the mean reference displacement magnitude
        and
        \subref{fig:experiments:invivo-results:mean-repe}
        the \glsxtrshort{mrepe}
        along the entire \gls{invivo} sequence for each method considered.
        In each \gls{bmode} image of the top row,
        the square region of interest is highlighted
        and the corresponding magnified inset displays the
        \ndim{2} displacement estimates.
        \Gls{bmode} images are displayed using a dynamic range of
        \SI{50}{\decibel}.
        The mean value (through time) of each quantity represented by a
        colorized solid line in
        \subref{fig:experiments:invivo-results:mean-disp-cpwc-87} and
        \subref{fig:experiments:invivo-results:mean-repe}
        is represented by a horizontal dashed line of the same color.
        These mean values were computed ignoring samples at \SI{2.2}{\second}
        due to the resulting extreme \glsxtrshort{mrepe} values.
        An animation of the figure and the corresponding slideshow
        are provided as supplementary material.
    }%
    \label{fig:experiments:invivo-results}
\end{figure*}

From the example images and corresponding displacement estimates
[\cref{%
    fig:experiments:invivo-results:pw,fig:experiments:invivo-results:cpwc-15,%
    fig:experiments:invivo-results:cnn,fig:experiments:invivo-results:cpwc-87%
}],
one can observe that \cpwcmethodname{1} suffered from diffraction artifacts
(mainly caused by \glspl{gl} and \glspl{sl} arising from the carotid walls),
disturbing both the speckle patterns and the resulting displacement estimates.
These artifacts were strongly reduced using \cpwcmethodname{15},
leading to speckle patterns similar to the reference ones (\cpwcmethodname{87}),
resulting in accurate displacement estimates.
The proposed \gls{cnn}-based imaging approach also reduced these artifacts,
restoring the underlying speckle patterns accurately.
This resulted in local displacement estimates with a quality similar to that
obtained with \cpwcmethodname{15}.

The analysis of the \gls{mrepe} values over time
[\cref{fig:experiments:invivo-results:mean-repe}]
shows that,
while \cpwcmethodname{1} was generally unable to estimate inter-frame
motion properly,
the proposed method resulted in high and stable displacement estimation quality,
similar to (though slightly worse than) \cpwcmethodname{15}.
This observation matches the results of the numerical experiments on
small displacements (\cref{sec:results:numerical-experiment}).
At \SI{2.2}{\second},
significant deviations in the \gls{mrepe} values for all methods compared
can be observed [\cref{fig:experiments:invivo-results:mean-repe}].
As the estimated reference tissue displacement
at this time instant is very small (\SI{\sim{} 5}{\micro\meter})
[\cref{fig:experiments:invivo-results:mean-disp-cpwc-87}],
local \gls{repe} values, and as a consequence \gls{mrepe} values,
can be very sensitive to small absolute errors.
Moreover,
it is likely that such small displacements are close to the minimum achievable
displacement estimation error (Cramér-Rao lower bound),
thus over amplifying the inherent sensitivity of the \gls{repe} to very small
reference displacements.
This behavior can also be observed in the numerical experiment
on small displacements towards the center of rotation of zones
\edtnumphtcylClbl{} and \edtnumphtcylDlbl{}
[\cref{fig:results:numerical-phantom:060}].
Therefore,
all values estimated at \SI{2.2}{\second}
were ignored in the computation of the following global metrics.
As global metrics,
we computed the mean value through time (ignoring said time instant)
of each estimated quantity
[represented as dashed lines in
\cref{fig:experiments:invivo-results:mean-disp-cpwc-87,%
fig:experiments:invivo-results:mean-repe}].
The estimated mean inter-frame displacement is \SI{66.83}{\micro\meter}.
The \gls{mrepe} values are \SIlist{90.18;13.38;20.73}{\percent} for
\cpwcmethodname{1}, \cpwcmethodname{15}, and the proposed \gls{cnn}-based
method, respectively.

\section{Discussion}%
\label{sec:discussion}

In this work,
we proposed a \ndim{2} motion estimation approach based on single unfocused
acquisitions to reconstruct consecutive frames and on pairs of consecutive
frames to estimate local displacements.
This approach relies on our \gls{cnn}-based image reconstruction
method~\cite{Perdios_ARXIV_2020a} to reconstruct full-view \gls{us} frames
from single unfocused acquisitions.
It consists of first reconstructing low-quality images using
a backprojection-inspired \gls{das} algorithm
and then feeding them to a \gls{cnn},
specifically trained to reduce diffraction artifacts inherent to
ultrafast \gls{us} imaging.
Inter-frame displacements are estimated by applying a state-of-the-art \ndim{2}
speckle tracking algorithm on consecutive frame pairs only.

\subsection{Performance in Numerical Conditions}%
\label{sec:discussion:numerical}

An important observation is that the proposed approach could not estimate
displacements accurately in zones dominated by \gls{ew} artifacts
(\cref{fig:results:loc-epe}, zone \edtnumphtcylBlbl{}).
This is directly related to the fact that the \gls{cnn} deployed
is not capable of restoring the underlying speckle patterns accurately
[\cref{fig:experiments:numerical-phantom:cnn}].
Slight improvements were observed compared with conventional single \gls{pw}
imaging (\cpwcmethodname{1}),
but far less striking than in zones dominated by \gls{sl} and \gls{gl} artifacts
(\cref{fig:results:loc-epe}, zones \edtnumphtcylClbl{} and \edtnumphtcylDlbl{}).
In~\cite{Perdios_ARXIV_2020a},
we already observed that \gls{ew} artifacts were the most difficult artifacts
to deal with,
but also that the restoration quality improved with the increase
of the \gls{cnn} capacity.
The latter implies that the reduction of these artifacts might be further
improved using a more efficient \gls{cnn} architecture or training process.

As expected,
we observed in the large-displacement case that
compounding multiple acquisitions in an attempt to improve
the obtained image quality induces strong motion artifacts,
mainly due to destructive interferences caused by axial motion.
In the presence of motion artifacts,
conventional \gls{cpwc}-based speckle tracking
was generally incapable of providing valid displacement estimation,
in particular in zones plagued by strong diffraction artifacts.
Consequently,
compounding multiple acquisitions decreased the displacement estimation quality
compared with single-\gls{pw} acquisitions (\cpwcmethodname{1}).
While motion compensation techniques have been proposed to tackle this
issue~\cite{Joos_UFFC_2018},
it remains unclear if motion-compensated coherent compounding
can be deployed in zones plagued by diffraction artifacts
(as it is based on inter-acquisition motion estimation),
and if it actually improves displacement estimation quality
in artifact-free zones compared with single unfocused acquisitions.
We demonstrated that the proposed single \gls{pw} \gls{cnn}-based approach is
capable of providing high-quality displacement estimates in artifact-free zones,
as well as in zones plagued by \gls{sl} and \gls{gl} artifacts.

In the case of small displacements,
increasing the number of compounded acquisitions using \gls{cpwc}-based tracking
progressively increased, as expected, the accuracy of displacement estimation.
The proposed \gls{cnn}-based approach achieves a displacement estimation quality
comparable to \cpwcmethodname{15} in zones suffering from \gls{sl} and \gls{gl}
artifacts and comparable to \cpwcmethodname{9} in artifact-free zones.
It can be noted that the relative estimation precision achieved by
the proposed approach was generally worse when analyzing
small inter-frame displacements than in larger displacement cases.
This was also observed for conventional \gls{cpwc}-based tracking
in artifact-free zones
[\gls{eg} compare \cpwcmethodname{1}, zone \edtnumphtcylAlbl{}
in \cref{fig:results:numerical-phantom:600,fig:results:numerical-phantom:060}].
This mainly comes from the fact that the minimum estimation error
of correlation-based tracking converges to a minimum value
(Cramér-Rao lower bound),
which, relatively speaking,
becomes more significant for smaller displacements~\cite{Pinton_UFFC_2006}.
For quantifying very small displacements,
applying speckle tracking to \gls{rf} data instead of envelope data
may improve precision~\cite{Walker_UFFC_1994},
\cite[Sec.~14.2.1]{Loizou_BOOK_2018},
at the expense of a reduced robustness to speckle decorrelation.

\subsection{Performance in Physical Conditions}%
\label{sec:discussion:invivo}

We demonstrated that the proposed \gls{cnn}-based approach,
which relies on single-\gls{pw} acquisitions,
significantly improved over conventional single \gls{pw} imaging
(\cpwcmethodname{1}).
It also achieved an accuracy of inter-frame displacement estimation
similar to that of \num{15} compounded acquisitions (\cpwcmethodname{15}),
in conditions where motion artifacts were negligible
and thus did not limit the performance of the comparative \cpwcmethodname{15}
method.

Overall,
the quantitative evaluations performed in the \gls{invivo} experiment
are comparable to those of the numerical one.
This does not only show that the proposed method can be applied
to \gls{invivo} data successfully,
even though the \gls{cnn} used for image reconstruction was trained
on simulated data only,
it also suggests that the results of the numerical experiments are
robust and translatable (to some extent) to experimental conditions.
More specifically,
as motion artifacts were negligible in the \gls{invivo} experiment,
the results obtained are best compared with the ones obtained
in the numerical experiment on small displacements
[\cref{fig:results:numerical-phantom:060}].
It can be noted that the artifacts initially shadowing the zone in which
displacement estimates were analyzed seem to be a combination of \gls{gl} and
\gls{sl} artifacts spawned by the highly echogenic carotid walls
[\cref{fig:experiments:invivo-results:pw}].
Thus,
zones \edtnumphtcylClbl{} and \edtnumphtcylDlbl{} of the numerical experiment
are of interest for comparison purposes,
as they contain \gls{sl} and \gls{gl} artifacts, respectively
[\cref{fig:experiments:numerical-phantom:pw}].
While the quantitative metrics are similar,
it is important to note that this presumptive combination of \gls{gl}
and \gls{sl} artifacts was not present in the numerical experiment,
and that the \enquote{signal-to-artifact} ratio was probably more favorable
in the \gls{invivo} experiment than in the numerical one.
One can observe that \cpwcmethodname{15} performed better than
the proposed method in the \gls{invivo} experiment
[\cref{fig:experiments:invivo-results:mean-repe}],
whereas both methods performed similarly well in zones \edtnumphtcylClbl{} and
\edtnumphtcylDlbl{} of the numerical experiment on small displacements
(\cref{tab:numerical-phantom-global-metrics}).
A performance drop of the proposed approach from numerical to physical
conditions was expected since the deployed \gls{cnn} was trained on simulated
images only.
This performance drop was already observed in~\cite{Perdios_ARXIV_2020a},
in which a detailed discussion on the discrepancies between the numerical
and physical conditions can be found.

It should be noted that the \gls{invivo} experiment was intentionally
carried out on a slow moving tissue zone.
This enabled us to obtain reference displacement estimates for
quantitative evaluation purposes,
and to select a frame rate, identical for all methods considered,
resulting in inter-frame displacements within ranges of interest.
However,
as speckle tracking is agnostic to the underlying frame rate,
the results are fully translatable to fast motion cases
with similar inter-frame displacement ranges,
provided that the required frame rate is achievable by the method deployed.

\subsection{Potential, Perspectives, and Limitations}%
\label{sec:discussion:potential}

The proposed approach is overall able to provide high-quality estimates
for a wide range of \ndim{2} inter-frame displacements,
even in tissue regions dominated by \gls{sl} and \gls{gl} artifacts.
As it only relies on single unfocused acquisitions to reconstruct consecutive
frames, it is immune to motion artifacts.
Moreover,
it is limited only by the propagation time of acoustic waves,
making it especially interesting for the analysis of rapidly changing events
at very high frames rates,
such as the propagation of shear waves in tissue
or complex flow patterns within the cardiovascular system,
where displacement estimation techniques based on multi-acquisition image
reconstruction methods may not be deployable.

The major limitation is that the current implementation of the proposed approach
was not able to provide accurate displacement estimates in regions
dominated by \gls{ew} artifacts.
This is most probably due to the fact that the patterns resulting from \gls{ew}
artifacts resemble speckle patterns much more closely than the ones
resulting from \gls{sl} and \gls{gl} artifacts
[\cref{fig:experiments:numerical-phantom:pw}].
Since \glspl{cnn} are,
in essence,
based on pattern recognition,
the close resemblance of two patterns,
one sought to be removed,
the other to be preserved,
represents a greater difficulty compared with a situation in which the two
patterns are very distinctive.
Both the \gls{ew} behavior and the general performance of the approach might
be further improved by augmenting the performance of the \gls{cnn} used
for image reconstruction.
For instance,
the use of a higher-capacity \gls{cnn} or a more efficient training process
may improve the restoration of tissue structures hidden by \gls{ew} artifacts.
Another way to tackle this limitation
would be to use transmit apodization~\cite{JensenJonas_UFFC_2016}.
This technique can significantly reduce \gls{ew} artifacts,
at the cost of limited energy towards the image borders.
However,
its effectiveness is limited by the apodization capability of \gls{us} systems,
in particular by the transmitter complexity.
If the method is not used at maximum achievable frame rate,
and in the presence of sufficiently stationary motion,
the robustness and precision of the displacement estimation could be improved,
for instance, by averaging multiple displacement estimates
or by using ensemble correlation~\cite{Perrot_IUS_2018}.

This study was limited to tracking fully developed speckle patterns,
hence no insights about tracking tissue structures arising from
specular or diffractive scattering should be drawn from it directly.
Yet,
carotid wall movement was observed to be similar to that of
conventional methods
(see animation of \cref{fig:experiments:invivo-results},
supplementary material).
The training set was also limited to simulated images
of fully developed speckle zones resulting from diffusive scattering.
In~\cite{Perdios_ARXIV_2020a},
we observed that,
while reconstructing other tissue structures is generally possible,
the performance may be less potent than in fully developed speckle zones.
Using a versatile training set may be considered to widen the
applicability of both the reconstruction approach and
the displacement tracking method proposed here.

On a more general perspective,
this work further validates the potency of the \gls{cnn}-based
image reconstruction method introduced in~\cite{Perdios_ARXIV_2020a}.
Indeed,
this method not only provides high-quality images
from single unfocused acquisitions,
but also preserves the information of underlying physical phenomena
that can be further exploited for estimating inter-frame displacements
accurately.

\section{Conclusion}%
\label{sec:conclusion}

In this work we proposed an approach for estimating
\ndim{2} inter-frame displacements in the context of ultrafast \gls{us} imaging.
The approach consists of a \gls{cnn} trained to restore high-quality images
from single unfocused acquisitions
and a speckle tracking algorithm to estimate inter-frame displacements
from two consecutive frames only.
Compared with conventional multi-acquisition strategies,
this approach is immune to motion artifacts
and enables accurate motion estimation at maximum frames rates,
even in highly heterogeneous tissues prone to strong diffraction artifacts.
Numerical and \gls{invivo} results demonstrated
that the proposed approach is capable of estimating displacement
vector fields from single-\gls{pw} acquisitions accurately,
including in zones initially hidden by \gls{sl} and \gls{gl} artifacts.
The proposed approach may thus unlock the full potential of
ultrafast \gls{us},
with direct applications to imaging modes that depend
on accurate motion estimation at maximum frame rates,
such as shear-wave elastography or ultrasensitive echocardiography.

\section*{Acknowledgment}%
\label{sec:acknowledgment}

The authors would like to thank Quentin Ligier for
his important contribution to the implementation
of the speckle tracking algorithm deployed in this work.
The authors would also like to thank the anonymous reviewers for their comments,
which contributed significantly to the improvement of this manuscript.

\bibliographystyle{IEEEtran}
\bibliography{resources/references}

\end{document}